\documentclass[twocolumn, floatfix, nofootinbib]{revtex4-1}

\usepackage{amsmath}
\usepackage{amsthm}
\usepackage{amsfonts}
\usepackage{mathtools}
\usepackage{dsfont}
\usepackage{amssymb}
\usepackage{bbold}
\usepackage{bm}
\usepackage{tensor}
\usepackage{physics}
\usepackage{graphicx}
\usepackage{empheq}
\usepackage{dsfont}
\usepackage{color}
\usepackage{booktabs}

\definecolor{niceBlue}{RGB}{20,10,237}
\usepackage[colorlinks,citecolor=niceBlue,linkcolor=niceBlue,urlcolor=niceBlue,hyperindex]{hyperref}

\newcommand{\lcm}{\operatorname{lcm}}
\newcommand{\pbceq}{\overset{\mathrm{pbc}}{=}}

\begin{document}

\title{Position-Dependent Excitations and UV/IR Mixing in the $\mathbb{Z}_{N}$ Rank-2 Toric Code\\
 and its Low-Energy Effective Field Theory
}

\author{Salvatore D. Pace}
\affiliation{Department of Physics, Massachusetts Institute of Technology, Cambridge, Massachusetts 02139, USA}

\author{Xiao-Gang Wen}
\affiliation{Department of Physics, Massachusetts Institute of Technology, Cambridge, Massachusetts 02139, USA}

\date{\today}

\begin{abstract}

We investigate how symmetry and topological order are coupled in the ${2+1}$d
$\mathbb{Z}_{N}$ rank-2 toric code for general $N$, which is an exactly
solvable point in the Higgs phase of a symmetric rank-2 $U(1)$ gauge theory.
The symmetry enriched topological order present has a non-trivial realization
of square-lattice translation (and rotation/reflection) symmetry, where anyons
on different lattice sites have different types and belong to different
superselection sectors. We call such particles ``position-dependent excitations.'' 
As a result, in the rank-2 toric code anyons can hop by one lattice site in some
directions while only by $N$ lattice sites in others, reminiscent of
fracton topological order in ${3+1}$d. We find that while there are
$N^2$ flavors of $e$ charges and $2N$ flavors of $m$ fluxes, there are not 
$N^{N^{2} + 2N}$ anyon types. Instead, there are $N^{6}$ anyon types, and we can use Chern-Simons theory with six $U(1)$ gauge fields to describe all of them. While the lattice translations permute anyon types, we find that such 
permutations cannot be expressed as transformations on the six $U(1)$ gauge 
fields. Thus the realization of translation symmetry in the $U^6(1)$ Chern-Simons 
theory is not known. Despite this, we find a way to calculate the translation-dependent properties of the theory. In particular, we find that the ground state
degeneracy on an ${L_{x}\times L_{y}}$ torus is ${N^{3}\gcd(L_{x},N)
\gcd(L_{y},N) \gcd(L_{x},L_{y},N)}$, where $\gcd$ stands for ``greatest common
divisor.'' We argue that this is a manifestation of UV/IR mixing which arises from the interplay between lattice symmetries and topological order.

\end{abstract}

\maketitle

\tableofcontents

\section{Introduction}\label{sec:intro}

Topological order~\cite{wen1990topological} is a cornerstone in understanding gapped liquid phases of
highly-entangled quantum matter~\cite{wen2019choreographed}. At the microscopic
level, phases with topological order exhibit long-range
entanglement~\cite{PhysRevLett.96.110404, PhysRevLett.96.110405,
PhysRevB.82.155138}. At the macroscopic level, from the highly-entangled
constituents emerges remarkable robust properties, like internal gauge fields,
exotic bulk excitations (anyons), topology-dependent ground state degeneracies,
and gapless chiral edge excitations. In the presence of symmetries, this
structure of topological orders becomes even richer and the quantum phase is
said to posses symmetry enriched topological (SET)
orders~\cite{PhysRevB.65.165113, PhysRevB.80.224406, PhysRevB.87.104406,PhysRevB.87.155115,PhysRevB.88.235103,teo2015theory,
PhysRevResearch.2.033331}.

For instance, the emergent anyons do not need to form a linear representation
of the symmetry group and can instead transform protectively under its
elements. This then allows them to carry fractional quantum numbers of the
symmetry, known as symmetry fractionalization~\cite{PhysRevB.65.165113,chen2017symmetry}.
This is a familiar phenomena in the context of fractional quantum Hall states
where the anyons transform protectively under the $U(1)$ symmetry group
corresponding to the electron's charge and consequentially carry fractional
amounts of the electron charge~\cite{PhysRevLett.50.1395}. Intrinsic
topological orders can also be enriched by external symmetries, such as the
space group of an underlying
lattice~\cite{PhysRevLett.90.016803,PhysRevB.74.174423,PhysRevB.90.121102,PhysRevB.96.195164}.
For example, anyons transforming protectively under lattice translations can
carry fractional crystal momentum, which consequentially reduces the size of
the first Brillouin zone in the reciprocal lattice~\cite{PhysRevB.65.165113}. In additional
to it's richness, we note how in both of the above examples, symmetry
fractionalization provides direct experimental signatures for the underlying
topological order: the former being fractionally quantized Hall conductivity in
a two-dimensional electron gas~\cite{PhysRevLett.48.1559} and the latter
through proposed neutron scattering experiments on candidate quantum spin
liquids in frustrated magnets~\cite{PhysRevB.90.121102,
PhysRevB.96.085136,PhysRevLett.121.077201}.

In both of the above examples, the symmetry elements act locally on anyons as a
$U(1)$ phases, which describe the fractional quantum number they carry.
However, it is also possible for the symmetry transformations to additionally
induce a nontrivial automorphism on the anyon types, permuting inequivalent
anyon types~\cite{PhysRevB.93.155121,Tarantino_2016,PhysRevB.100.115147}. By
inequivalent anyons, we mean excitations belonging to different topological
superselection sectors (see appendix E of Ref.~\cite{kitaev2006anyons} for a
review of the algebraic theory of anyons). For instance, consider a
double-layer fractional quantum Hall system. There is an internal $Z_{2}$
symmetry operator which exchanges the elementary anyons on each layer,
physically corresponding to an anyon tunneling between
layers~\cite{PhysRevB.81.045323}. Furthermore, there are exactly solvable
lattice models, like Wen's plaquette model~\cite{PhysRevB.78.155134,
PhysRevB.86.161107} or the color
code~\cite{PhysRevLett.97.180501,PhysRevB.90.115118}, where lattice
transformations permute inequivalent anyons. For example, in the plaquette
model on the square lattice, there are two types of elementary excitations, $e$
charges and $m$ flux. Lattice translations by one lattice space in either the
$x$ or $y$ directions take all $e$ charges to $m$ fluxes and all $m$ fluxes to
$e$ charges. Following the terminology introduced in
Ref.~\cite{PhysRevB.93.155121}, we say that SET phases including such
nontrivial automorphisms contain unconventional SET orders. In these phases,
the interplay between symmetry and topological order is even more striking.
Indeed, in all of the above described examples, the existence of automorphism
permuting inequivalent anyons causes the topological ground state degeneracy to
become dependent on the system's size (in a non-extensive way).

While gauging a global symmetry leads to topological order present in the
discussion above, gauging a subsystem symmetry (a symmetry acting only on
subspace of the entire system) leads to fracton topological
order~\cite{PhysRevB.92.235136,
PhysRevB.94.155128,10.21468/SciPostPhys.6.4.041}. Phases with this order are
characterized by topological excitations that are only able to move along
subsets of the spatial lattice~\cite{PhysRevLett.94.040402, PhysRevA.83.042330,
PhysRevB.88.125122, nandkishore2019fractons, pretko2020fracton}. These
subdimensional excitations are also said to have fractionalized mobility. While
individually they can only move within planes (planons), along lines (lineons),
or are entirely immobile (fractons), their composite objects can be completely
mobile. Such subdimensional physics has emerged as a very exciting frontier of
quantum matter, displaying a wide range of phenomena, such as nonergodic
behavior~\cite{PhysRevLett.94.040402, PhysRevB.95.155133, PhysRevX.9.021003}
and emergent gravitational
physics~\cite{PhysRevB.74.224433,gu2012emergence,PhysRevD.81.104033,PhysRevD.96.024051,
PhysRevB.99.155126, PhysRevB.100.245138,PhysRevB.102.161119}

There's an interesting way to understand subdimensional particles' mobility
based on how their topological superselection sectors transform under lattice
translations~\cite{PhysRevB.100.195136}. Consider a superselection sector $s$
and the operator $T_{i}$ which performs a translation in the $i$-direction by
one lattice spacing. If lattice translations induce the transformation ${T_{i}:
s\mapsto s}$, then by definition there exists an operator that moves the
elementary excitation of superselection sector $s$ by one lattice spacing in
the $i$ direction (a ``string'' operator). However, suppose that for all
integers $n$ less than the linear size of the system, that ${(T_{i})^{n}:
s\mapsto s_{n}\neq s}$. This then means that there does not exist a string
operator which moves the elementary excitation of $s$ in the direction $i$. For
instance, on a cubic lattice if ${(T_{x})^{n}: s\mapsto s_{n}\neq s}$ and
${(T_{y})^{n}: s\mapsto s_{n}\neq s}$ but ${T_{z}: s\mapsto s}$, then $s$ is
the topological superselection sector of a lineon restricted to move in the $z$
direction. If ${(T_{i})^{n}: s\mapsto s_{n}\neq s}$ for all $i$ then $s$
describes a fracton, where as if ${T_{i}: s\mapsto s}$ for all $i$ then $s$
describes a normal mobile excitation.

This is reminiscent of unconventional SET orders.  Indeed, under translations
in any direction a subdimensional particle cannot move, its corresponding
superselection sector is changed.  Therefore, for phases with fracton
topological order and an underlying lattice, there is always some lattice
translation that induces a nontrivial automorphism on the superselection
sectors.  This point of view is quite enlightening.  Indeed, it provides an
intuitive explanation as to why the topological ground state degeneracy scales
with system size for fracton topological orders: the number of excitation types
grows with the system size.  Furthermore, the UV/IR mixing known to occur in
fracton phases can be understood as a consequence of global equivalence
relations between excitation types which only exist when the system is put on a
topologically nontrivial space.  The intuition behind this will be explained in
detail throughout this paper.  Therefore, from this point of view, some of the
most striking features of fracton topological phases arises from a rich
interplay between long-range entanglement and symmetry.

Throughout this paper, we'll denote excitations that change type under lattice
transformations as ``position-dependent excitations.'' From the above
discussion, all subdimensional excitations are position-dependent, but not all
position-dependent excitations are subdimensional.  Indeed, position-dependent
excitations occur in the aforementioned Wen's plaquette model and the color
code, which both have traditional topological order.  Because they do not
posses fracton topological order, it means that while there are superselection
sectors where ${T_{i}: s\mapsto s'\neq s}$, there is some $n>1$ smaller than
the linear system size such that ${(T_{i})^{n}: s\mapsto s}$.  Recalling the
example provided from the plaquette model, while a single translation takes
${T_{i}:e\mapsto m}$ and ${T_{i}:m \mapsto e}$, a double translation acts as
the identity: ${(T_{i})^{2}:e\mapsto e}$ and ${(T_{i})^{2}:m \mapsto m}$.

In this manuscript, we investigate an unconventional SET phase with
position-dependent excitations that are closer in similarity to subdimensional
excitations.  In other words, an unconventional SET order that is similar to
fracton topological order.  One promising route to such an SET order is to
start off with a phase containing only fracton topological order and undergo a
phase transition to a phase with conventional topological order.  Indeed, when
subdimensional particles condense if all of the additional excited particles
that usually prevent their movement are absorbed into the condensate, they can
become mobile~\cite{PhysRevB.96.195139, PhysRevB.97.235112, PhysRevB.98.035111,
PhysRevB.104.165121, arxiv.2207.00409}.  And so, through condensing excitations, an extensive
subset of the superselection sectors (a feature of fracton
topological order) reduce down to to a finite number (conventional topological
order).

A particularly simple class of theories known to include subdimensional
particles are symmetric $U(1)$ tensor gauge theories~\cite{PhysRevB.95.115139}.
In the context of quantum matter, these tensor gauge theories are effective
theories describing exotic quantum spin liquid phases~\cite{PhysRevB.74.224433,
gu2012emergence, PhysRevLett.124.127203, PhysRevB.105.L060408}. Unlike in
fracton topological order, the low-energy physics is governed by gapless
excitations and the gauge charges' subdimensional nature arises due to emergent
higher-moment symmetries, like dipole momentum conservation. When these tensor
gauge theories are Higgsed such that the $U(1)$ group reduces to
$\mathbb{Z}_{N}$, the gapless gauge boson becomes gapped, the subdimensional
particles typically become mobile, and the phase posses topological
order~\cite{PhysRevB.96.195139, PhysRevB.97.235112,
PhysRevB.98.035111,oh2021path, oh2022effective}. While these excitations have become mobile,
they can typically only hop by multiple sites at a time in the directions they
were previously immobile. For example, a lineon belonging to superselection
sector $s$ in a phase where it is partially condensed still satisfies ${T_{x}:
s\mapsto s'\neq s}$, ${T_{y}: s\mapsto s'\neq s}$ and ${T_{z}: s\mapsto s}$,
like before the phase transition, but now also satisfies ${(T_{x})^{n}:
s\mapsto s}$ and ${(T_{y})^{n}: s\mapsto s}$ for an integer $n>1$ less than the
system size. Therefore, this Higgs transition has induced a phase transition
from gapless fracton order to unconventional SET order. This SET phase is more
``subdimensional like'' because the value of $n$ grows with $N$, where as, for
instance, the plaquette model always had $n=2$.

The remaining of this paper is organized as follows. In
Section~\ref{sec:review}, we start by reviewing the rank-2 $U(1)$ ``scalar
charge'' gauge theory in ${2+1}$d and its subdimensional particles, and describe how to
regulate it on a spatial square lattice. Upon Higgsing the lattice gauge fields
of this tensor gauge theory, we introduce the exactly solvable model studied
throughout this paper: the $\mathbb{Z}_{N}$ rank-2 toric code. In
Section~\ref{sec:posDepExci}, we show how emergent conservation laws of gauge
charge and flux arising from the fusion rules enforce that
anyons of the same species having a hidden flavor index and carry different charge/flux based on their position.
This recovers previous results of the mobility of these excitations and their
position-dependent braiding statistics. Additionally, it allows us to define
the anyon lattice which reveals how the excitations change type under lattice
transformations, hence making the rank-2 Toric code poses an unconventional SET
order. Interestingly, these automorphisms on anyon lattice vectors are
nonlinear. Then, in Section~\ref{sec:globalEquiv}, we consider the affect of
periodic boundary conditions and find new equivalence relations between anyon types which arise from
the lattice-translations' realization on the anyon lattice. We find the ground state degeneracy for general $N$ and identify new
non-local string operators that further modifies the mobility of the
excitations. The ground state degeneracy sensitively depends on the system size,
which we discuss in the context of UV/IR mixing. The position-dependent
excitation picture and the anyon lattice additional provides a straight forward way for developing a
mutual Chern-Simons theory for the rank-2 Toric Code, which is the subject of
Section~\ref{sec:CSthy}. Using the Chern-Simon gauge fields, we find a basis
set of holonomies for the torus in Section~\ref{sec:holonomies}, which in
Section~\ref{sec:effectiveAction} we use to find a low-energy effective action
in terms of the gauge fields' zero modes. The number of ground states from this
effective action is the same as that found by considering the anyon lattice
group. Furthermore, this low-energy effective action explicitly depends on the
number of unit cells from the microscopic theory, revealing the origin of the
UV/IR mixing in the effective theory.

\section{$\mathbb{Z}_{N}$ Rank-2 Toric Code in $2+1$d}\label{sec:bigR2TC}

One of the simplest cases of topological order is $\mathbb{Z}_{N}$ topological
order. A canonical system which contains $\mathbb{Z}_{N}$ topological order is
Kitaev's toric code model\footnote{Throughout this paper, the terminology
``toric code'' refers to the general $\mathbb{Z}_{N}$ version} in $2+1$d
spacetime dimensions~\cite{kitaev2003fault}. It is an exactly solvable model
that resides in the deconfined phase of a $\mathbb{Z}_{N}$ quantum gauge
theory. One way to motivate the toric code Hamiltonian is by ``Higgsing'' a
vector $U(1)$ lattice gauge theory down to a $\mathbb{Z}_{N}$ theory. The
$U(1)$ gauge group is reduced to $\mathbb{Z}_{N}$, causing the $U(1)$ gauge
charge to become a $\mathbb{Z}_{N}$ gauge charge, the $U(1)$ magnetic flux loop
to become a $\mathbb{Z}_{N}$ vortex, and the gapless gauge boson (the photon)
to become gapped~\cite{PhysRevD.19.3682}. The toric code is an exactly solvable
point in this Higgs phase of the $U(1)$ lattice gauge theory.

In this section, we investigate the rank-2 toric code, a recent generalization
of the toric code. We will first review how its Hamiltonian can be obtained by
Higgsing the gauge field in a tensor $U(1)$ lattice gauge
theory~\cite{PhysRevB.97.235112, PhysRevB.98.035111,oh2021path, oh2022effective}. We then
introduce a position-dependent excitation picture, from which we study the
excitations' mobility, braiding statistics, symmetry properties and find the
ground state degeneracy for general $N$. In addition to it's utility in this
section, the anyon lattice framework for the position-dependent excitations
will be crucial for developing a mutual Chern-Simons theory of the rank-2 toric
code in Section~\ref{sec:CSthy}.

\subsection{Higgsed $U(1)$ Symmetric Tensor Gauge Theory and its
Excitations}\label{sec:review}

\subsubsection{Continuum Field Theory}\label{sec:U1TGT}

Consider a rank-2 $U(1)$ quantum gauge theory in the continuum with a compact
gauge field $A^{ij}(x)$ and conjugate electric field $E^{ij}(x)$. We work in
$2+1$d so the indices $i,j\in\{x,y\}$. Both of these are symmetric rank-2
quantum tensor fields and satisfy the canonical commutation relation
\begin{equation}\label{eqn:continuumCanonicalCom}
    \left[A^{ij}(x), E^{kl}\left(y\right)\right]=\mathrm{i}\left(\delta\indices{^{ki}} \delta\indices{^{lj}} +\delta\indices{^{li}}  \delta\indices{^{kj}} \right) \delta\left(x-y\right).
\end{equation}
We work consider the so-called scalar charge theory, where the $U(1)$ gauge charge density $\rho$ is given by the Gauss's law
\begin{equation}\label{eqn:gLaw}
\rho(\bm{x}) = \partial_{i}\partial_{j}E^{ij}(\bm{x}),
\end{equation}
where Einstein's summation convention is assumed and ${\partial_{i}\equiv \partial/\partial x^{i}}$. Another thoroughly studied symmetric tensor gauge theory is the so-called vector charge theory, where the gauge charge density is a vector field and satisfies the Gauss's law is ${\rho^{i} = \partial_{j}E^{ji}}$~\cite{PhysRevB.95.115139}, but in ${2+1}$d the scalar and vector charge theories are dual to one another~\cite{oh2021path}. The Gauss's law in Eq.~\eqref{eqn:gLaw} generates the gauge transformation~\cite{rasmussen2016stable}
\begin{equation}\label{eqn:gTrans}
A^{ij}(\bm{x}) \to A^{ij}(\bm{x}) + \partial_{i}\partial_{j}f(\bm{x}),
\end{equation}
where $f(\bm{x})$ is a general function. In light of this gauge transformation, we define the components of the gauge-invariant magnetic field as\footnote{In the 3+1d scalar charge theory, the components of the magnetic field $\tilde{B}^{ij}$ are defined as $\tilde{B}^{ij} = \epsilon^{iab}\partial_{a}A^{bj}$, where $\epsilon^{ijk}$ is the totally antisymmetric Levi-Civita symbol. In the 2+1d theory, we define the components of the vector magnetic field $\bm{B}$ in terms of $\tilde{B}$ as $B^{x} = -\tilde{B}^{zy}$ and $B^{y} = \tilde{B}^{zx}$, which leads to Eq.~\eqref{eqn:magneticFieldComps}.}
\begin{subequations}\label{eqn:magneticFieldComps}
\begin{align}
    B^{x}(\bm{x}) &= \partial_{y}A\indices{^{xy}}(\bm{x}) - \partial_{x}A\indices{^{yy}}(\bm{x}),\\
    B^{y}(\bm{x}) &= \partial_{x}A\indices{^{xy}}(\bm{x}) - \partial_{y}A\indices{^{xx}}(\bm{x}).
\end{align}
\end{subequations}
We note that in this form, given that $A$ transforms like a 2-tensor, the magnetic field $\bm{B}$ transforms as a vector.

Symmetric tensor gauge theories have attracted a vast interests recently due to their matter excitations having subdimensional mobility due to global conservation laws~\cite{PhysRevB.95.115139}. For instance, the $U(1)$ dipole moment $\bm{x}\rho$ is conserved:
\begin{equation}\label{eqn:dipCons}
    \int x^{i}\rho = \int x^{i}\partial_{j}\partial_{k}E^{jk} = \text{b.t.} -\int \partial_{k}E^{ik}  = \text{b.t.},
\end{equation}
where ``b.t.'' stands for ``boundary term.'' Therefore, allowed dynamical processes are only those that conserve the system's dipole moment. So, an isolated gauge charge cannot move while a two particle bound state can. By itself, it is immobile and hence a fracton. Additionally, the ``magnetic angular momentum,'' ${(xB^{y}-yB^{x})}$, is conserved:
\begin{equation}\label{eqn:rCrossBcons}
\begin{aligned}
\int (xB^{y}-yB^{x}) &= \int (x\partial_{x}A\indices{^{xy}} - x\partial_{y}A\indices{^{xx}},\\
& \hspace{50pt}-y\partial_{y}A\indices{^{xy}} + y \partial_{x}A\indices{^{yy}}),\\
&= \int (-A\indices{^{xy}}+A\indices{^{xy}}) + \text{b.t.} = \text{b.t.}.
\end{aligned}
\end{equation}
This is like the ``angular momentum'' conservation law for the vector charge theory, which enforces vector gauge charges to move only in the direction of their charge. Therefore, $U(1)$ magnetic flux loops can move only in the direction of $\bm{B}$ and are therefore lineons.

\subsubsection{Lattice Field Theory and Higgsing}\label{sec:R2TC}

To regularize the continuum theory on a lattice, we discretized the two-dimensional space as a square lattice while time remains a continuous variable. Throughout this section, there will be objects acting on or residing on the sites, edges, and plaquette of the square lattice. To make the notation less cumbersome, we'll label all of these by a corresponding lattice site. For each lattice site $(x,y)$, we associate with it the plaquette whose center is at ${(x+\frac{1}{2},y + \frac{1}{2})}$, the horizontal edge whose center is ${(x+\frac{1}{2},y)}$, and the vertical edge whose center is at ${(x, y+\frac{1}{2})}$. Throughout this paper, lengths are measured in units of the lattice constant and so the lattice constant is unity.

In order to discretize the continuum tensor gauge fields, first consider two $U(1)$ quantum rotors residing on each lattice site and one $U(1)$ quantum rotor at each plaquette. Because the 2-tensor fields are symmetric in two spatial dimensions, they each contain three independent operators. For a given lattice site, the operators corresponding to the $xx$ and $yy$ components each act on one of the rotors residing on the lattice site while the operator corresponding to the $xy$ component acts on the rotor residing on the plaquette~\cite{xu2006novel}. Therefore, the lattice operator $A^{xx}_{x,y}$ acts one one of the rotors at lattice site $(x,y)$, $A^{yy}_{x,y}$ acts on the other rotor on the lattice site, and $A^{xy}_{x,y}$ acts on the rotor on the plaquette corresponding to $(x,y)$. This can be motivated from the gauge transformation in Eq.~\eqref{eqn:gTrans} as $A^{ij}$ should act on the same location as $\partial_{i}\partial_{j}$. The designations then follow directly from the fact that the discretized differential operators ${\partial_{x}\partial_{x}}$ and ${\partial_{y}\partial_{y}}$ are naturally associated with a lattice site while ${\partial_{x}\partial_{y}}$ is naturally associated with a plaquette~\cite{PhysRevB.96.035119}.

In the continuum theory, the canonical commutation relation Eq.~\eqref{eqn:continuumCanonicalCom} is manifestly symmetric in exchanging the indices of $A$ or $E$. However, because of the Kronecker delta functions, the components of the tensor fields satisfy ${[A^{xx}_{x,y},E^{xx}_{x,y}] = 2i}$, ${[A^{yy}_{x,y},E^{yy}_{x,y}] = 2i}$, and ${[A^{xy}_{x,y},E^{xy}_{x,y}] = i}$. Because $A^{ij}$ is compact, this implies that while $E^{xy}$ has integer eigenvalues, $E^{xx}$ and $E^{yy}$ have only even integer eigenvalues. Following Ref.~\cite{PhysRevB.98.035111}, we make the transformation $E^{xx}\to 2E^{xx}$ and $E^{xyy}\to 2E^{yy}$ so the lattice variables all satisfy ${[A^{ij}_{x,y},E^{ij}_{x',y'}] = \mathrm{i}\delta_{x,x'}\delta_{y,y'}}$ and thus all components $E^{ij}$ have integer eigenvalues. With this change, however, from the discretized Gauss's law the eigenvalues of the $\rho$ operator take only even integer eigenvalue. Therefore, we also make the transformation ${\rho \to 2\rho}$. Then, the eigenvalues of $E^{ij}$ and $\rho$ are only integers and the lattice Gauss's law becomes
\begin{equation}
\begin{aligned}
\rho_{x,y} &\hspace{-1pt}=\hspace{-1pt} E_{x+1,y}^{x x} \hspace{-1pt}+ E_{x-1,y}^{x x}\hspace{-1pt}-2 E_{x,y}^{x x}\hspace{-1pt}+E_{x,y+1}^{y y}\hspace{-1pt} +E_{x,y-1}^{y y} \\
&\hspace{10pt} -\hspace{-1pt}2 E_{x,y}^{y y}\hspace{-1pt}+\hspace{-1pt}E_{x,y}^{x y}\hspace{-1pt}-\hspace{-1pt}E_{x-1,y}^{x y}\hspace{-2pt}+\hspace{-1pt}E_{x-1,y-1}^{x y}\hspace{-2pt}-\hspace{-1pt}E_{x,y-1}^{x y}.
\end{aligned}
\end{equation}
Additionally, the components of the lattice magnetic field are given by
\begin{subequations}
\begin{align}
B_{x,y}^{x} & = (A_{x,y}^{x y}-A_{x,y-1}^{x y}) - ( A_{x+1,y}^{y y} - 
A_{x,y}^{y y} ),\\
B_{x,y}^{y} & =   (A_{x,y}^{x y} - A_{x-1,y}^{x y}) - ( A_{x,y+1}^{x x} - A_{x,y}^{x x}).
\end{align}
\end{subequations}

\begin{figure}[t!]
\centering
    \includegraphics[width=.48\textwidth]{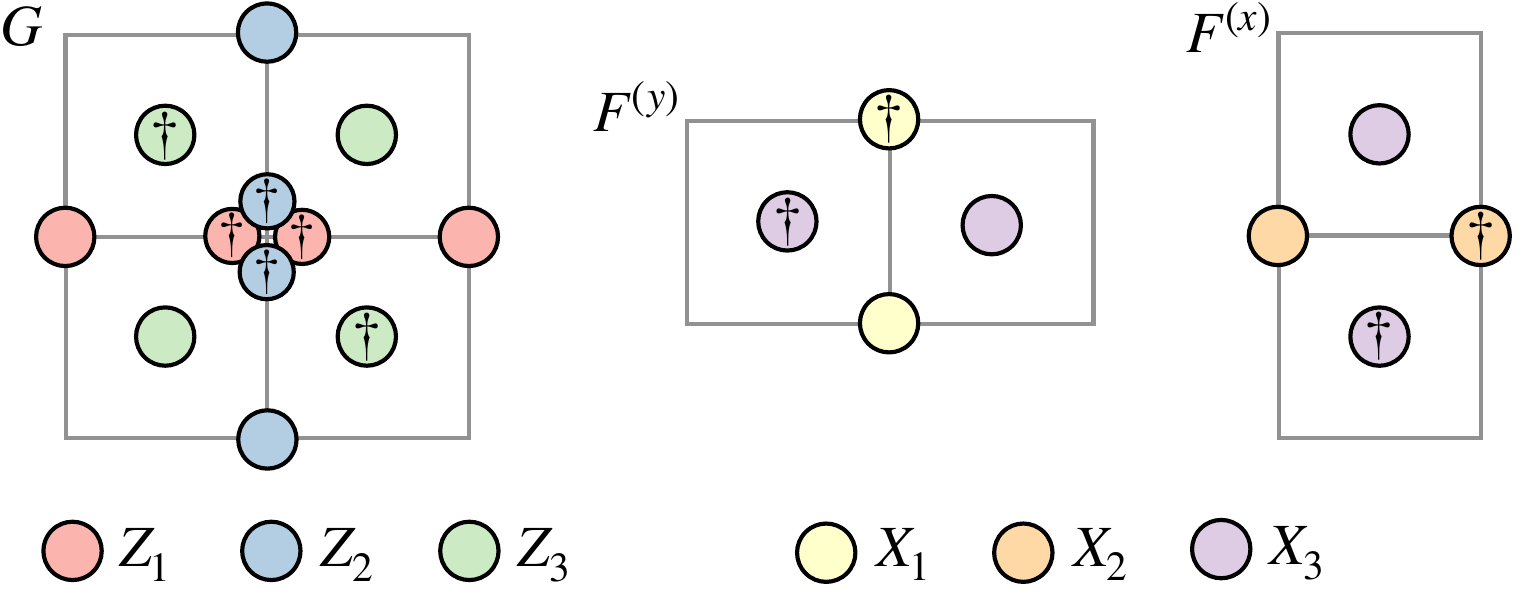}
    \caption{%
    A graphical representation of the $\mathbb{Z}_{N}$ Gauss operator $G$ and magnetic flux operators $F^{(x)}$ and $F^{(y)}$ contained in the rank-2 toric code's Hamiltonian (see Eq.~\eqref{eqn:Hamiltonian}). The disks are color-coded to represent $X_{i}$ and $Z_{i}$ operators, according to the legend. Furthermore, disks with a $\dagger$ represent the Hermitian conjugate of the corresponding operator.
    }
    \label{fig:ops}
\end{figure}

Having put the scalar charge $U(1)$ gauge theory onto a lattice, The $U(1)$ gauge group is now Higgsed so all charge-$N$ excitations condense into the vacuum, reducing the $U(1)$ gauge group down to $\mathbb{Z}_{N}$~\cite{PhysRevD.19.3682}. This can be done formally by introducing a charge-$N$ matter field, including a Higgs term in the $U(1)$ lattice gauge theory Hamiltonian, and then considering the low-energy subspace in the Higgs phase where the gauge field is constrained to ${A^{ij} = 2\pi(\text{integer})/N}$~\cite{PhysRevB.97.235112, PhysRevB.98.035111}. In this Higgs phase, there are $\mathbb{Z}_{N}$ lattice gauge fields $X_{i}$ and $\mathbb{Z}_{N}$ electric fields $Z_{i}$ which are given by
\begin{subequations}
\begin{align}
    X_{1} &=  \mathrm{e}^{\mathrm{i}A^{xx}},
    \quad\quad
    X_{2} =  \mathrm{e}^{\mathrm{i}A^{yy}},%
    \quad\quad
    X_{3} =  \mathrm{e}^{\mathrm{i}A^{xy}},\label{eq:Xdefinition}\\
    Z_{1} &=  \omega^{E^{xx}},
    \quad\quad
    Z_{2} =  \omega^{E^{yy}},
    \quad\quad
    Z_{3} =  \omega^{E^{xy}},
\end{align}
\end{subequations}
where $\omega = \mathrm{e}^{2\pi \mathrm{i}/N}$. It follows that $X_{i}$ and $Z_{j}$ are unitary operators and satisfy ${Z_{j}X_{i} = \omega^{\delta_{i,j}}X_{i}Z_{j}}$, $X_{i}^{N} = 1$, and $Z_{i}^{N} = 1$. Additionally, there is a $\mathbb{Z}_{N}$ Gauss operator $G_{x,y}$ and $\mathbb{Z}_{N}$ magnetic flux operators $F^{(x)}_{x,y}$ and $F^{(y)}_{x,y}$ that are given by
\begin{subequations}\label{eqn:ZnGaussLawandMagFlux}
\begin{align}
    G_{x,y} &= \omega^{\rho_{i}},\\
    F^{(x)}_{x,y} &=  e^{\mathrm{i}B^{x}_{x,y}},\\
    F^{(y)}_{x,y} &=  e^{\mathrm{i}B^{y}_{x,y}}.
\end{align}
\end{subequations}
In terms of the $X_{i}$ and $Z_{i}$ operators, $G_{x,y}$, $F^{(x)}_{x,y}$, and $F^{(y)}_{x,y}$ are
\begin{subequations}
\begin{align}
    G_{x,y} &= (Z^{\dagger}_{1,x,y})^{2}(Z^{\dagger}_{2,x,y})^{2}Z_{3,x,y}Z^{\dagger}_{3,x-1,y}Z_{3,x-1,y-1}\nonumber\\
    &\hspace{10pt}\times Z^{\dagger}_{3,x,y-1}Z_{1,x-1,y}Z_{1,x+1,y}Z_{2,x,y-1}Z_{2,x,y+1},\\
    F^{(x)}_{x,y} &=  X_{2,x,y}X^{\dagger}_{2,x+1,y}X_{3,x,y}X^{\dagger}_{3,x,y-1},\\
    F^{(y)}_{x,y} &=  X_{1,x,y}X^{\dagger}_{1,x,y+1}X_{3,x,y}X^{\dagger}_{3,x-1,y}.
\end{align}
\end{subequations}
A graphical representation of these operators is shown in Fig.~\ref{fig:ops}.

Using these operators, the rank-2 toric code Hamiltonian is
\begin{equation}\label{eqn:Hamiltonian}
H_{\text{R2TC}} = -\frac{1}{2}\sum_{x,y}(G_{x,y} + F^{(x)}_{x,y} + F^{(y)}_{x,y} + \text{h.c.}).
\end{equation}
It is straight-forward to confirm that $G$, $F^{(x)}$, and $F^{(y)}$ are all mutually commuting for every lattice site. Therefore, this model is exactly solvable and the ground state $\ket{\text{vac}}$ is the eigenstate of $G$,  $F^{(x)}$, and $F^{(y)}$ with the maximum eigenvalue, which is 1.

Before moving on to discuss the excitations in this model, we now consider lattice transformations and show that $H$ is invariant under the space group of the square lattice. The space group is $p4m$ and can be generated by a 4-fold rotation about a lattice site $C_{4}$, a mirror reflection about a horizontal line that intersects lattice sites $M_{x}$, and lattice translations. It is easy to see that $H_{\text{R2TC}}$ is invariant under translations. To see that it is invariant under the point group elements, first note that in the $U(1)$ theory, because $A^{ij}$ is a symmetric tensor its components transform under $C_{4}$ as ${A^{xx}_{x,y}\to A^{yy}_{-y,x}}$, ${A^{yy}_{x,y}\to A^{xx}_{-y,x}}$, and ${A^{xy}_{x,y}\to -A^{xy}_{-y-1,x}}$, and under $M_{x}$ as ${A^{xx}_{x,y}\to A^{xx}_{x,-y}}$, ${A^{yy}_{x,y}\to A^{yy}_{x,-y}}$, and ${A^{xy}_{x,y}\to -A^{xy}_{x,-y-1}}$. Therefore, according to Eq.~\eqref{eq:Xdefinition}, the $X_{i}$ operators transform as
\begin{subequations}
\begin{align}
    &\quad X_{1,x,y}\to X_{2,-y,x},\nonumber\\
    C_{4}:&\quad X_{2,x,y}\to X_{1,-y,x},\\
    &\quad X_{3,x,y}\to  X^{\dagger}_{3,-y-1,x},\nonumber\\\nonumber\\
    &\quad X_{1,x,y}\to X_{1,x,-y},\nonumber\\
    M_{x}:&\quad X_{2,x,y}\to X_{2,x,-y},\\
    &\quad X_{3,x,y}\to  X^{\dagger}_{3,x,-y-1}.\nonumber
\end{align}
\end{subequations}
Because $E_{ij}$ is also a symmetric tensor, $Z_{i}$ transforms in the same way as $X_{i}$. Therefore, $G$, $F^{(x)}$, and $F^{(y)}$ transform like
\begin{subequations}
\begin{align}
    &\quad G_{x,y}\to G_{-y,x},\nonumber\\
    C_{4}:&\quad F^{(x)}_{x,y}\to F_{-y,x}^{y},\label{eqn:c4OppTrans}\\
    &\quad F^{(y)}_{x,y}\to (F^{(x)}_{-y-1,x})^{\dagger},\nonumber\\\nonumber\\
    &\quad G_{x,y}\to G_{x,-y},\nonumber\\
    M_{x}:&\quad F^{(x)}_{x,y}\to F^{(x)}_{x,-y},\label{eqn:MxOppTrans}\\
    &\quad F^{(y)}_{x,y}\to (F^{(y)}_{x,-y-1})^{\dagger}.\nonumber
\end{align}
\end{subequations}
Using this, it is easy to see that the Hamiltonian remains unchanged by both $C_{4}$ and $M_{x}$ and therefore it is invariant under all lattice transformation.

\begin{figure}[t!]
\centering
    \includegraphics[width=.48\textwidth]{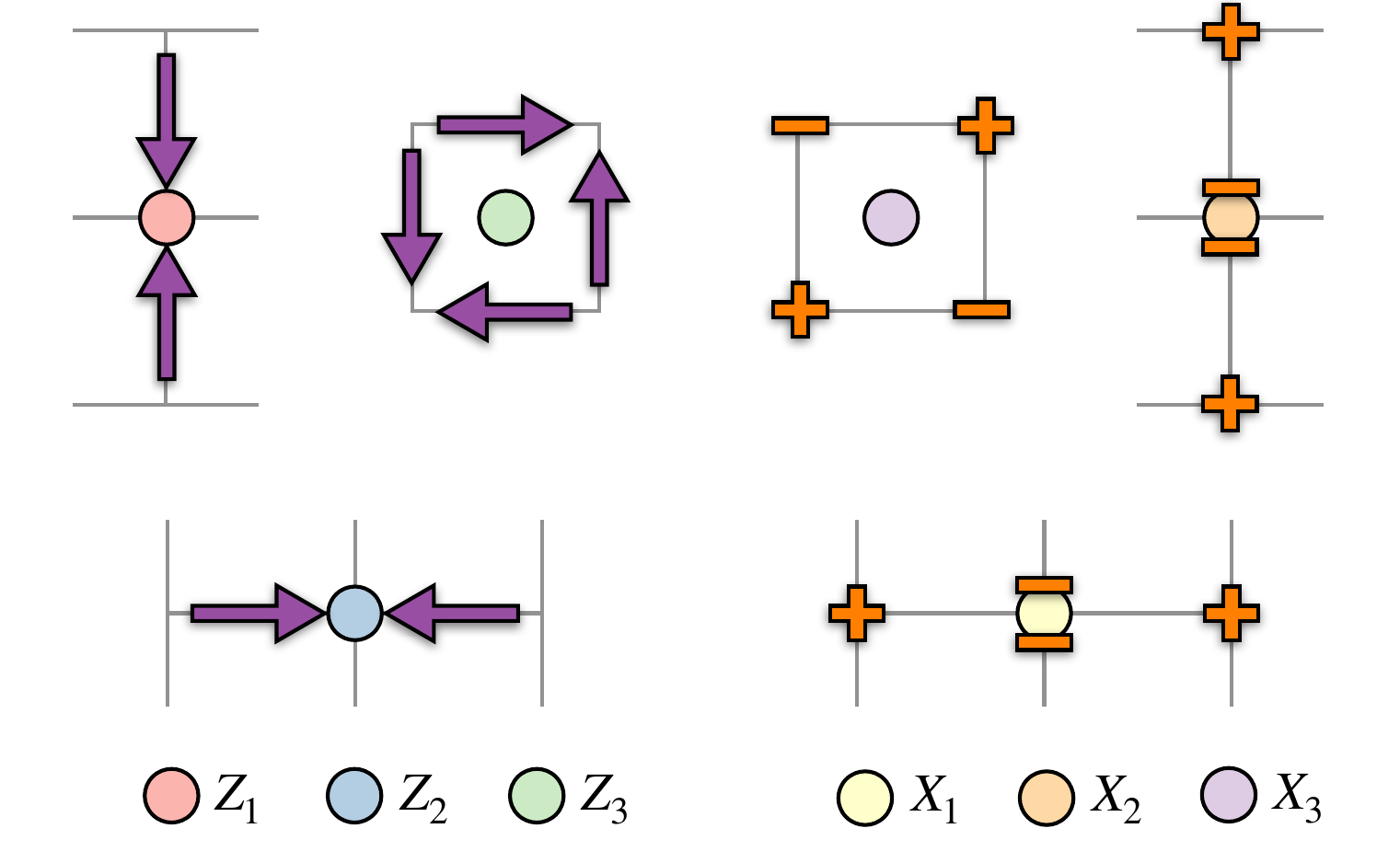}
    \caption{%
    Excitations in the rank-2 toric code can be excited using the $X_{i}$ and $Z_{i}$ operators and carry charge as defined by Eq.~\eqref{eqn:ChargeDef}.
    (Left) The $Z_{i}$ operators excite vector gauge fluxes $\vec{m} = (m^{(x)},m^{(y)})$. $m^{(x)}$ ($m^{(y)}$) excitations reside on horizontal (vertical) links, and a positive gauge flux pictorially corresponds to a vector pointing in the $+x$ ($+y$) direction.
    (Right) The $X_{i}$ operators excite gauge charges $e$, which we represent as ``$+$'' and ``$-$'' signs on the lattice sites for positive and negative charge, respectively.
    }
    \label{fig:excitations}
\end{figure}

The many-body ground state satisfies the local constraints ${G_{x,y} = F^{(x)}_{x,y} = F^{(y)}_{x,y} = 1}$ at each lattice site. The excited states are connected to the ground state by acting $X_{i}$ or $Z_{i}$ on $\ket{\text{vac}}$, which gives rise to violations of the ground state constraints. Just like in the toric code model, this corresponds to exciting gapped particles from an artificial vacuum. Violations of the of the ${G_{x,y}=1}$ constraint corresponds to exciting gauge charges, which we'll denote as $e$ particles, whereas violations of the ${F^{(x)}_{x,y}=1}$ and ${F^{(y)}_{x,y}=1}$ constrains correspond to exciting gauge fluxes (vortices), which we denote as $m^{(x)}$ and $m^{(y)}$ particles, respectively. We'll denote the charge carried by an excitation as the mathfrak font of the symbol used to label the excitation. So, the gauge charge carried by $e_{x,y}$ (an $e$ particle at lattice site $(x,y)~$) is $\mathfrak{e}_{x,y}$ and the gauge flux carried by $m^{(x)}_{x,y}$ is $\mathfrak{m}^{(x)}_{x,y}$ and by $m^{(y)}_{x,y}$ is $\mathfrak{m}^{(y)}_{x,y}$. If it appears strange to label the charge by the lattice site it corresponds to, we note that this will be discussed in greater detail throughout Section~\ref{sec:anyonLattice}. Nonetheless, for a general energy eigenstate $\ket{\psi}$, the amount of charge carried by an excitation at lattice site $(x,y)$ is determined by the eigenvalue relations
\begin{subequations}
    \label{eqn:ChargeDef}
    \begin{align}
        G_{x,y}\ket{\psi} &= \omega^{\mathfrak{e}_{x,y}}\ket{\psi},\\
        F^{(x)}_{x,y}\ket{\psi} &= \omega^{\mathfrak{m}^{(x)}_{x,y}}\ket{\psi},\\
        F^{(y)}_{x,y}\ket{\psi} &= \omega^{\mathfrak{m}^{(y)}_{x,y}}\ket{\psi}.
    \end{align}
\end{subequations}
The symmetry properties of the charges can then be determined from Eqs.~\eqref{eqn:c4OppTrans} and~\eqref{eqn:MxOppTrans}. Indeed, transforming $G_{x,y}$, $F^{(x)}_{x,y}$, and $F^{(y)}_{x,y}$ reveal that  $\mathfrak{e}_{x,y}$, $\mathfrak{m}^{(x)}_{x,y}$, and $\mathfrak{m}^{(y)}_{x,y}$ transform under $C_{4}$ and $M_{x}$ as
\begin{subequations}\label{eqn:excitationsLatticeTrans}
\begin{align}
    &\quad \mathfrak{e}_{x,y}\to \mathfrak{e}_{-y,x},\nonumber\\
    C_{4}:&\quad \mathfrak{m}^{(x)}_{x,y}\to \mathfrak{m}_{-y,x}^{(y)},\label{eqn:c4transLatticeCharge}\\
    &\quad \mathfrak{m}^{(y)}_{x,y}\to -\mathfrak{m}^{(x)}_{-y-1,x},\nonumber\\\nonumber\\
    &\quad \mathfrak{e}_{x,y}\to \mathfrak{e}_{x,-y},\nonumber\\
    M_{x}:&\quad \mathfrak{m}^{(x)}_{x,y}\to \mathfrak{m}^{(x)}_{x,-y},\\
    &\quad \mathfrak{m}^{(y)}_{x,y}\to -\mathfrak{m}^{(y)}_{x,-y-1}.\nonumber
\end{align}
\end{subequations}

The $e_{x,y}$ excitations naturally reside on lattice sites and, like in the continuum theory before Higgsing, are scalar gauge charges. However, $m^{(x)}$ and $m^{(y)}$ transform into each other like the components of a vector. Because of this, we introduce the vector charge ${\vec{m}_{x,y} = (m^{(x)}_{x,y},m^{(y)}_{x,y})}$, and consider $m^{(x)}_{x,y}$ ($m^{(t)}_{x,y}$) to reside on the horizontal (vertical) edge associated with the lattice site $(x,y)$\footnote{Unlike the normal toric code, due to $e$ excitations being particles while $\vec{m}$ excitations are vectors, the rank-2 toric code lacks an electric-magnetic duality.}. As components of a vector charge, pictorially we represent them as vectors, where a positive charge $m^{(x)}$ ($m^{(y)}$) is a vector pointing in the $+x$ ($+y$) direction. The six different types of elementary excitations excited using $X_{i}$ and $Z_{i}$ operators are shown in Fig.~\ref{fig:excitations}.

\subsection{Position-Dependent Excitations}\label{sec:posDepExci}

In this section, instead of using the lattice operators $X_{i}$ and $Z_{i}$ in detail to study the excitations of the rank-2 toric code, we'll instead consider the anyon lattice they form. This will present a powerful picture where anyons of the same species (for instance, both $e$ excitations) carry different gauge charge depending on their position. This turns out to be a natural framework to understand their mobility and braiding statistics. It also reveals how the anyons couple with the lattice symmetries in a rich way. Finally, this will act as a starting point for developing a Chern-Simons theory as a low-energy effective theory for the rank-2 toric code in Section~\ref{sec:CSthy}. We note that throughout this section, we will \textit{not} be assuming periodic boundary conditions. Instead, their affect will be investigate in Section~\ref{sec:globalEquiv}.

\subsubsection{The Anyon Lattice}\label{sec:anyonLattice}

In the rank-2 toric code, as was discussed Section~\ref{sec:R2TC}, there are three species of elementary excitations: $e$ and ${\vec{m} = (m^{(x)}, m^{(y)})}$. It is typically the case that excitations of the same species carry the same gauge charge/flux, and so for each species of anyon there is only one inequivalent anyon flavor. However, this is not generally true and the number of inequivalent elementary excitations can be greater than the number of species.

The number of inequivalent elementary excitations can be found by first considering the most general possibility where for every lattice site the $e$ and $\vec{m}$ particles carry different gauge charges and fluxes. Therefore, for the rank-2 toric code this initial starting point is when the gauge charges and fluxes satisfy $\mathfrak{e}_{x_{1},y_{1}}\neq \mathfrak{e}_{x_{2},y_{2}}$ and $\vec{\mathfrak{m}}_{x_{1},y_{1}}\neq \vec{\mathfrak{m}}_{x_{2},y_{2}}$ for ${(x_{1},y_{1})\neq(x_{2},y_{2})}$. This means that for an $L_{x}\times L_{y}$ size system, there are initially $3L_{x}L_{y}$ inequivalent elementary excitations. By requiring that gauge charge and flux is locally conserved by all processes that excite $e$ and $\vec{m}$ excitations, subsets of the initial gauge charges/flux $\mathfrak{e}_{x,y}$, $\mathfrak{m}^{(x)}_{x,y}$, and $\mathfrak{m}^{(y)}_{x,y}$ will become linearly dependent. As a consequence, when all such equivalence relations are taken into account the initial general $3L_{x}L_{y}$ inequivalent elementary excitations reduces to the actual number of inequivalent elementary excitations.

This procedure is general and can be used for any topological order provided its fusion rules. For instance, in the $2+1$d toric code, the initial $2L_{x}L_{y}$ number of elementary excitations reduces to two: a single $\mathbb{Z}_{N}$ gauge charge and $\mathbb{Z}_{N}$ gauge flux. And so, as is already known in the $2+1$d toric code, all charges and vortices carry the same gauge charge and gauge flux, respectively, regardless of their position on the lattice. However, for the rank-2 toric code we'll find that this is no longer the case.

There are six different ways to excite $e$ and $\vec{m}$ particles from the ground state in the rank-2 toric code, which are the configurations shown in Fig.~\ref{fig:excitations}. All other ways to locally excite excitations are combinations of these six configurations. They can be translated into fusion rules. Letting $\mathbf{1}$ denote the trivial excitation (the vacuum), for every lattice site $(x,y)$ there are three fusion rules involving gauge fluxes
\begin{subequations}\label{eqn:mFusionRules}
\begin{align}
    \mathbf{1} &= m^{(x)}_{x-1,y}\otimes \bar{m}^{(x)}_{x,y},\label{eqn:mFusionRules1}\\
    \mathbf{1} &= m^{(y)}_{x,y-1}\otimes \bar{m}^{(y)}_{x,y},\label{eqn:mFusionRules2}\\
    \mathbf{1} &= \bar{m}^{(x)}_{x,y}\otimes \bar{m}^{(y)}_{x,y} \otimes m^{(x)}_{x,y+1}\otimes m^{(y)}_{x+1,y}\label{eqn:mFusionRules3},
\end{align}
\end{subequations}
and three fusion rules involving gauge charges
\begin{subequations}\label{eqn:eFusionRules}
\begin{align}
    \mathbf{1} &= e_{x-1,y} \otimes \bar{e}_{x,y} \otimes \bar{e}_{x,y} \otimes e_{x+1,y},\label{eqn:eFusionRules1}\\
    \mathbf{1} &= e_{x,y-1} \otimes \bar{e}_{x,y}\otimes \bar{e}_{x,y} \otimes e_{x,y+1},\label{eqn:eFusionRules2}\\
    \mathbf{1} &= e_{x,y}\otimes \bar{e}_{x+1,y}\otimes e_{x+1,y+1}\otimes \bar{e}_{x,y+1}\label{eqn:eFusionRules3}.
\end{align}
\end{subequations}
Here, we use the notation that, for instance, $\bar{e}$ denotes the anti-particle of $e$ and so, by definition, they obey the fusion rules ${\mathbf{1} = \bar{e}_{x,y}\otimes e_{x,y}}$, ${\mathbf{1} = \bar{m}^{(x)}_{x,y}\otimes m^{(x)}_{x,y}}$, and ${\mathbf{1} = \bar{m}^{(y)}_{x,y}\otimes m^{(y)}_{x,y}}$. These fusion rules define equivalence relations between excitations. Particularly, that the composite objects on the right-hand side Eqs.~\eqref{eqn:mFusionRules} and~\eqref{eqn:eFusionRules} belong to the same topological superselection sector as the trivial excitation $\mathbf{1}$. Instead of thinking about the superselection sectors, we can equally view these as an emergent conservation laws relating the charge and flux carried by different excitations. Because the ground state carries no charge and flux, these fusion rules therefore give
\begin{subequations}\label{eqn:gaugeConLaws}
\begin{align}
    \mathfrak{m}^{(x)}_{x-1,y} - \mathfrak{m}^{(x)}_{x,y} &= 0,\label{eqn:gaugeFluxCons1}\\
    \mathfrak{m}^{(y)}_{x,y-1} - \mathfrak{m}^{(y)}_{x,y} &= 0,\label{eqn:gaugeFluxCons2}\\
    -\mathfrak{m}^{(x)}_{x,y} - \mathfrak{m}^{(y)}_{x,y} + \mathfrak{m}^{(x)}_{x,y+1} + \mathfrak{m}^{(y)}_{x+1,y} &= 0,\label{eqn:gaugeFluxCons3}\\
    \mathfrak{e}_{x-1,y} - 2~\mathfrak{e}_{x,y} + \mathfrak{e}_{x+1,y} &= 0,\label{eqn:gaugeChargeCons1}\\
    \mathfrak{e}_{x,y-1} - 2~\mathfrak{e}_{x,y} + \mathfrak{e}_{x,y+1} &= 0,\label{eqn:gaugeChargeCons2}\\
    \mathfrak{e}_{x,y} - \mathfrak{e}_{x+1,y} + \mathfrak{e}_{x+1,y+1} - \mathfrak{e}_{x,y+1} &= 0.\label{eqn:gaugeChargeCons3}
\end{align}
\end{subequations}
In what follows, we'll treat these as recurrence relations and recursively solve for the gauge charge and flux carried by $e$, $m^{(x)}$, and $m^{(y)}$ at a general lattice site\footnote{Alternatively, one could view Eq.~\eqref{eqn:gaugeConLaws} as finite differences, which in the continuum limit become the differential equations ${\partial_{x}m^{(x)} = 0}$, ${\partial_{y}m^{(y)} = 0}$, and ${\partial_{y}m^{(x)} + \partial_{x}m^{(y)} = 0}$ for the gauge fluxes and ${\partial_{x}^{2}e = 0}$, ${\partial_{y}^{2}e = 0}$, and ${\partial_{x}\partial_{y}e = 0}$ for the gauge charge. Then, it is clear that their position-dependency are ${m^{(x)}(y) = C_{1} + y~C_{2}}$, ${m^{(y)}(x) = C_{3} - x~C_{2}}$, and ${e(x,y) = C_{4} + x~C_{5} + y~C_{6}}$, where $C_{i}$ are all constants. This is exactly what we find solving these recursively.}.

Let's first consider the equivalence relations Eqs.~\eqref{eqn:gaugeFluxCons1} and~\eqref{eqn:gaugeFluxCons2}. They imply that for every fixed value $y$ that
\begin{equation}\label{eqn:mxLongDir}
    \mathfrak{m}^{(x)}_{x_{1},y} = \mathfrak{m}^{(x)}_{x_{2},y}\quad\forall~x_{1}~\text{and}~x_{2},
\end{equation}
and that for every fixed value $x$ that
\begin{equation}\label{eqn:myLongDir}
    \mathfrak{m}^{(y)}_{x,y_{1}} = \mathfrak{m}^{(y)}_{x,y_{2}}\quad\forall~y_{1}~\text{and}~y_{2}.
\end{equation}
Therefore, all $m^{(x)}$ along the same horizontal line, or similarly all $m^{(y)}$ along the same vertical line, carry the same gauge flux. So, for determining the number of inequivalent elementary excitations, we can restrict ourselves to only having to consider gauge flux types $\mathfrak{m}^{(x)}_{0,y}$ and $\mathfrak{m}^{(y)}_{x,0}$ since $\mathfrak{m}^{(x)}_{x,y} = \mathfrak{m}^{(x)}_{0,y}$ for all $x$ and $\mathfrak{m}^{(y)}_{x,y} = \mathfrak{m}^{(y)}_{x,0}$ for all $y$. Using this, the remaining equivalence relation for the gauge fluxes, Eq.~\eqref{eqn:gaugeFluxCons3}, becomes
\begin{equation*}
     -\mathfrak{m}^{(x)}_{0,y} - \mathfrak{m}^{(y)}_{x,0} + \mathfrak{m}^{(x)}_{0,y+1} + \mathfrak{m}^{(y)}_{x+1,0} = 0.
\end{equation*}
Solving for $\mathfrak{m}^{(x)}_{0,y+1}$ and setting $x=0$ gives a recurrence relation for $\mathfrak{m}^{(x)}_{0,y}$ in terms of $\mathfrak{m}^{(x)}_{0,y-1}$ and the $y$-independent $\mathfrak{m}^{(y)}_{0,0}$ and $\mathfrak{m}^{(y)}_{1,0}$. Recursively solving this for $\mathfrak{m}^{(x)}_{0,y}$ gives
\begin{equation*}
    \mathfrak{m}^{(x)}_{1,y} = y~\mathfrak{m}_{0,0}^{(y)} +  \mathfrak{m}_{0,0}^{(x)}  -y~ \mathfrak{m}^{(y)}_{1,0}.
\end{equation*} 
Similarly, solving instead for $\mathfrak{m}^{(y)}_{x+1,0}$ and setting $y=0$ gives a recurrence relation for $\mathfrak{m}^{(y)}_{x,0}$ in terms of $\mathfrak{m}^{(y)}_{x-1,0}$ and the $x$-independent $\mathfrak{m}^{(x)}_{0,0}$ and $\mathfrak{m}^{(x)}_{0,1}$. Recursively solving this for $\mathfrak{m}^{(y)}_{x,0}$ gives
\begin{equation*}
    \mathfrak{m}^{(y)}_{x,0} = x~\mathfrak{m}^{(x)}_{0,0} + \mathfrak{m}^{(y)}_{0,0} -x~ \mathfrak{m}_{0,1}^{(x)}.
\end{equation*}
These expressions are in terms of $\mathfrak{m}_{0,0}^{(x)}$, $\mathfrak{m}_{0,0}^{(y)}$, $\mathfrak{m}_{0,1}^{(x)}$, and $\mathfrak{m}_{1,0}^{(y)}$, which are not linearly independent as they're related to one another by Eq.~\eqref{eqn:gaugeFluxCons3} with ${x=y=0}$. Expressing $\mathfrak{m}_{1,0}^{(y)}$ in terms of the other three, we are left with only $\mathfrak{m}_{0,0}^{(x)}$, $\mathfrak{m}_{0,0}^{(y)}$, and $\mathfrak{m}_{0,1}^{(x)}$. It's convenient to introduce the gauge fluxes
\begin{equation}\label{eq:basisgaugeflux}
    \mathfrak{m}^{x} = \mathfrak{m}^{(x)}_{0,0},
    \quad\quad
    \mathfrak{m}^{y} = \mathfrak{m}_{0,0}^{(y)},
    \quad\quad
    \mathfrak{g} = \mathfrak{m}^{(x)}_{0,1} - \mathfrak{m}^{(x)}_{0,0},
\end{equation}
and then the above results yield that the gauge flux carried by the excitations $m^{(x)}$ and $m^{(y)}$ associated with lattice site $(x,y)$ are
\begin{subequations}\label{eqn:mChargeLatDep}
\begin{align}
    \mathfrak{m}^{(x)}_{x,y} &= \mathfrak{m}^{x} + y~ \mathfrak{g},\\
    \mathfrak{m}^{(y)}_{x,y} &= \mathfrak{m}^{y} -x~ \mathfrak{g}.
\end{align}
\end{subequations}

The same type of recursive analysis can be done for the $e$ particles. Indeed, the equivalence relation provided by Eq.~\eqref{eqn:gaugeChargeCons1} give a recurrence relation for $\mathfrak{e}_{x,y}$ in terms of $\mathfrak{e}_{x+1,y}$ and $\mathfrak{e}_{x+2,y}$. Similarly, Eq.~\eqref{eqn:gaugeChargeCons2} give a recurrence relation for $\mathfrak{e}_{x,y}$ in terms of $\mathfrak{e}_{x,y+1}$ and $\mathfrak{e}_{x,y+2}$. Solving these two recurrence relation independently give
\begin{align*}
    \mathfrak{e}_{x,y} &= x~\mathfrak{e}_{1,y} + (1-x)~\mathfrak{e}_{0,y},\\
    \mathfrak{e}_{x,y} &= y~\mathfrak{e}_{x,1} +(1-y)~\mathfrak{e}_{x,0}.
\end{align*}
Plugging one of these into the other yields an expression for $\mathfrak{e}_{x,y}$ in terms of $\mathfrak{e}_{0,0}$, $\mathfrak{e}_{0,1}$, $\mathfrak{e}_{1,0}$, and $\mathfrak{e}_{1,1}$. However, these four gauge charges are linearly dependent, related to another another by Eq.~\eqref{eqn:gaugeChargeCons3} at ${x=y=0}$. Expressing $\mathfrak{e}_{1,1}$ in terms of the other three and introducing the gauge charges
\begin{equation}\label{eq:basisgaugecharge}
    \mathfrak{e} = \mathfrak{e}_{0,0},
    \quad\quad
    \mathfrak{p}^{x} = \mathfrak{e}_{1,0} - \mathfrak{e}_{0,0},
    \quad\quad
    \mathfrak{p}^{y} = \mathfrak{e}_{0,1} - \mathfrak{e}_{0,0},
\end{equation}
the expression for $\mathfrak{e}_{x,y}$ simplifies to
\begin{equation}\label{eqn:eChargeLatDep}
    \mathfrak{e}_{x,y} = \mathfrak{e}+ x~\mathfrak{p}^{x} + y~\mathfrak{p}^{y}.
\end{equation}

\begin{figure}[t!]
\centering
    \includegraphics[width=.48\textwidth]{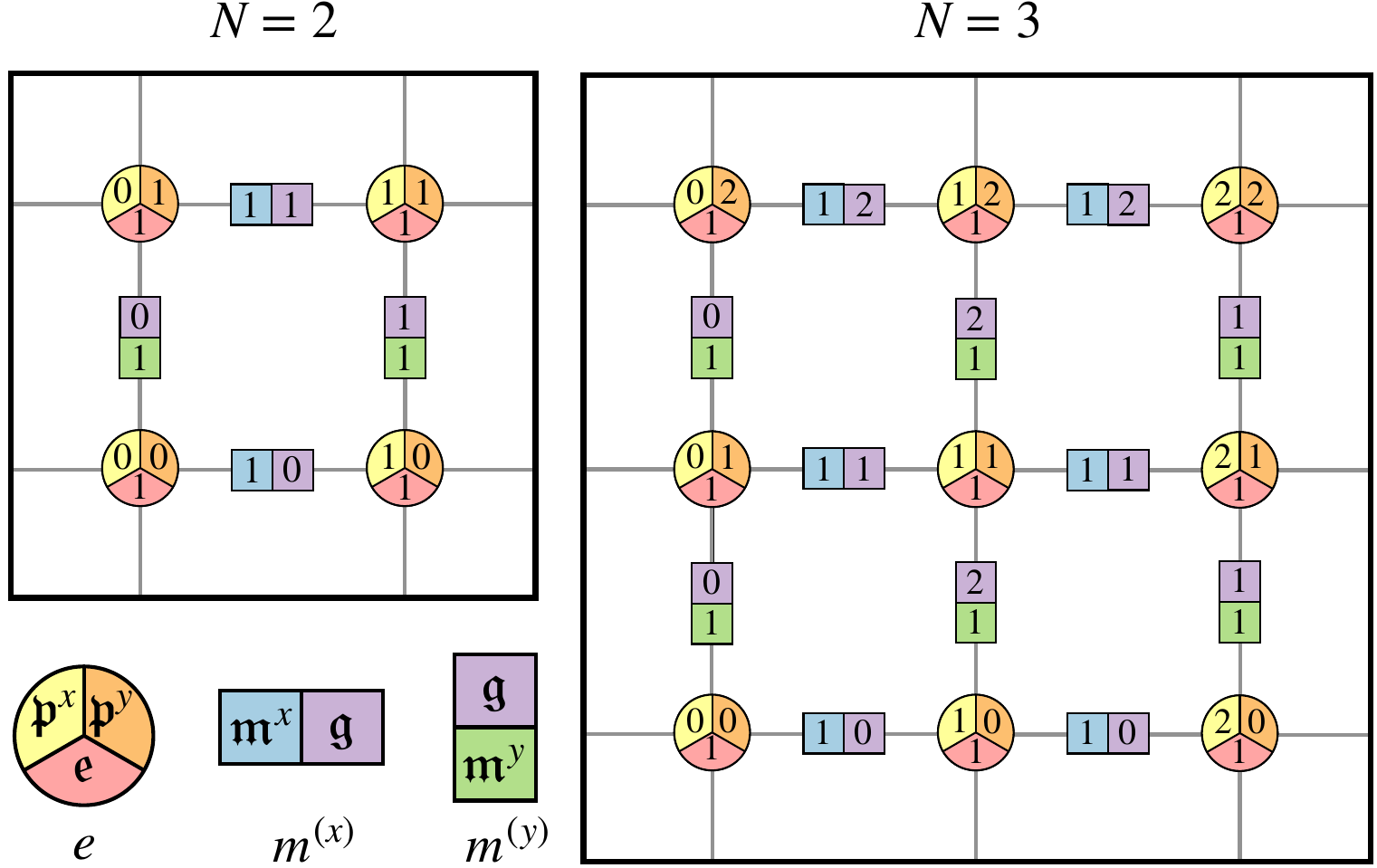}
    \caption{%
    The gauge charge and gauge flux carried by excitations in the rank-2 toric code depend on the particle's position (see Eqs.~\eqref{eqn:mChargeLatDep} and~\eqref{eqn:eChargeLatDep}). For the $\mathbb{Z}_{N}$ theory, this causes the unit cell of the spatial lattice to become size $N\times N$. Here, we show examples of the unit cell for when (left) $N=2$ and when (right) $N=3$.
    The three types of gauge charges carried by $e$ particles ($\mathfrak{e},~\mathfrak{p}^{x},~\mathfrak{p}^{y}$) are graphically represented by the color-coded thirds of the circles on each lattice site.
    Similarly, the three gauge flux types $\vec{m}$ particles carry ($\mathfrak{m}^{x},~\mathfrak{m}^{y},~\mathfrak{g}$) are graphically represented by the color-coded halves of the rectangles on the links of the lattice.
    The integers labeling both represent the number basis gauge charges and fluxes a single elementary excitation at that location carries.
    }
    \label{fig:AnyonUnitCell}
\end{figure}

At first glance, Eqs.~\eqref{eqn:mChargeLatDep} and~\eqref{eqn:eChargeLatDep} appear to imply that all excitations at different lattice sites carry different gauge charges/fluxes, which would imply that there is an extensive number of anyons flavors. For a theory with the fusion rules Eqs.~\eqref{eqn:mFusionRules} and~\eqref{eqn:eFusionRules} where $e$ and $\vec{m}$ are, for instance, $U(1)$ gauge charges/fluxes, this is indeed true. However, this is not the case for the rank-2 toric code because the $e$ and $\vec{m}$ particles carry $\mathbb{Z}_{N}$ charges and fluxes, respectively, and therefore obey the fusion rules ${\mathbf{1} = (e_{x,y})^{N}}$, ${\mathbf{1} = (m^{(x)}_{x,y})^{N}}$, and ${\mathbf{1} = (m^{(y)}_{x,y})^{N}}$. In terms of the basis charges and fluxes, this means that they all satisfy ${N~\mathfrak{m}^{x} = 0}$, ${N~\mathfrak{m}^{y} = 0}$, etc. Therefore, Eqs.~\eqref{eqn:mChargeLatDep} and~\eqref{eqn:eChargeLatDep} satisfy
\begin{subequations}\label{eqn:chargeNconstraint}
\begin{align}
    \mathfrak{e}_{x,y} &= \mathfrak{e}_{x+N,y} = \mathfrak{e}_{x,y+N},\label{eqn:chargeNconstraintePart}\\
    \mathfrak{m}^{(x)}_{x,y} &= \mathfrak{m}^{(x)}_{x,y+N},\label{eqn:mxNconstraintePart}\\
    \mathfrak{m}^{(y)}_{x,y} &= \mathfrak{m}^{(y)}_{x+N,y},\label{eqn:myNconstraintePart}
\end{align}
\end{subequations}
and so the number of anyon flavors is independent of system size. There are $N$ flavors of $m^{(x)}$ particles, $N$ flavors of $m^{(y)}$ particles, and $N^{2}$ flavors of $e$ particles, each carrying different combinations of gauge flux and gauge charge, respectively. This causes the unit cell of the square lattice to become enlarged, now being $N\times N$ lattice sites in size. Fig.~\ref{fig:AnyonUnitCell} shows examples of the unit cell for $N=2$ and $N=3$ with the lattice sites and edges labeled by the gauge charge/flux that an excitation there carries.

From the above analysis, the $m^{(x)}$ and $m^{(y)}$ excitations carry three different types of gauge flux ($\mathfrak{m}^{x}$,~$\mathfrak{m}^{y}$,~and~$\mathfrak{g}$) while the $e$ excitations carry three different types of gauge charge ($\mathfrak{e}$, $\mathfrak{p}^{x}$, and $\mathfrak{p}^{y}$). Using this, we can now introduce the anyon lattice that describes the excitations. For abelian anyons, the anyon lattice $\mathcal{A}$ is an abelian group under fusion. Anyons can be represented as vectors and the fusion of anyons correspond to vector addition. In this representation, the basis vectors spanning the anyon lattice correspond to the basis gauge fluxes and charges. Every vector ${\vec{\ell}\in\mathcal{A}}$ corresponds to a unique topological superselection sector. Therefore, for the rank-2 toric code $\mathcal{A} = \mathbb{Z}^{6}_{N}$, and a generic excitation is represented by the anyon lattice vector $\vec{\ell}$ as
\begin{equation}\label{eqn:genericExcitationOnAnyonLattice}
\begin{aligned}
\vec{\ell} &= \ell_{1}~\vec{\mathfrak{m}^{x}} + \ell_{2}~\vec{\mathfrak{m}^{y}} + \ell_{3}~\vec{\mathfrak{g}} + \ell_{4}~\vec{\mathfrak{e}} +
    \ell_{5}~\vec{\mathfrak{p}^{x}} + 
    \ell_{6}~\vec{\mathfrak{p}^{y}},\\
    &\doteq
    \begin{pmatrix}
         \ell_{1} \\
         \ell_{2} \\
         \ell_{3}
    \end{pmatrix}
    \oplus
    \begin{pmatrix}
         \ell_{4} \\
         \ell_{5} \\
         \ell_{6}
    \end{pmatrix},
\end{aligned}
\end{equation}
It's important to note that because $\ell_{i}\in\mathbb{Z}_{N}$, the elementary excitations represented as anyon lattice vector are
\begin{align}
    \vec{\ell}_{m^{(x)}_{x,y}} &= \begin{pmatrix}
         1 \\
         0 \\
         y~\operatorname{mod}~N
    \end{pmatrix}
    \oplus
    \begin{pmatrix}
         0 \\
         0 \\
         0
    \end{pmatrix},\\[0.5em]
    \vec{\ell}_{m^{(y)}_{x,y}} &= \begin{pmatrix}
         0 \\
         1 \\
         -x~\operatorname{mod}~N
    \end{pmatrix}
    \oplus
    \begin{pmatrix}
         0 \\
         0 \\
         0
    \end{pmatrix},\\[0.5em]
    \vec{\ell}_{e_{x,y}} &= \begin{pmatrix}
         0 \\
         0 \\
         0
    \end{pmatrix}
    \oplus
    \begin{pmatrix}
         1 \\
         x~\operatorname{mod}~N \\
         y~\operatorname{mod}~N
    \end{pmatrix}
\end{align}

Finally, we note that because the anyon lattice is spanned by three $\mathbb{Z}_{N}$ gauge charges and three $\mathbb{Z}_{N}$ gauge fluxes, the rank-2 toric code in $2+1$d possesses $\mathbb{Z}^{3}_{N}$ topological order~\cite{PhysRevB.97.235112, PhysRevB.98.035111}. However, we emphasize that because it is $\mathbb{Z}^{3}_{N}$ topological order in the presence of the square lattice symmetries, this is really symmetry enriched topological (SET) order. As we'll see in Section~\ref{sec:latticeTransonAnyonLat} and throughout later parts of the paper, this plays an important role in understanding the rank-2 toric code. For instance, the lattice symmetry elements permute the anyon flavors, causing the topological degeneracy to be extremely sensitive to the system's size.

\subsubsection{Pseudo-Subdimensional Particles}\label{sec:mobility}

In general, an excitation can only reside on lattice sites/edges that are compatible with the gauge charge/flux it carries. Consequentially, an excitation's mobility can be affected if the gauge charge/flux it carries depends on its position. For instance, an $e$ excitation at $(x_{1},y_{1})$ can only move to the lattice site $(x_{2},y_{2})$ if ${\mathfrak{e}_{x_{1},y_{1}} = \mathfrak{e}_{x_{2},y_{2}}}$ or else local conservation of gauge charge will be violated. Because this emergent conservation law arises from the fusion rules, this equivalently means that in order for a local operator to exist that hops an excitation from $(x_{1},y_{1})$ to $(x_{2},y_{2})$, then it must be the case that  ${\mathfrak{e}_{x_{1},y_{1}} = \mathfrak{e}_{x_{2},y_{2}}}$. Here, we consider the mobility of $e$, $m^{(x)}$, and $m^{(y)}$ excitations using the position-dependent anyon-lattice picture, compare them to the continuum field theory of Section~\ref{sec:U1TGT} where they correspond to subdimensional particles, and discuss their string operators found in Refs.~\cite{PhysRevB.97.235112, PhysRevB.98.035111, oh2021path}.

First, consider the $m^{(x)}$ and $m^{(y)}$ excitations. Let's denote the direction orthogonal (parallel) to their superscript as the transverse (longitudinal) direction. So, for instance, the transverse (longitudinal) direction for $m^{(x)}$ is the $y$ ($x$) direction. The mobility of $m^{(x)}$ and $m^{(y)}$ in their longitudinal directions is determined by Eqs.~\eqref{eqn:mxLongDir} and ~\eqref{eqn:myLongDir}, respectively. Because all edges in their longitudinal direction are associated with the same gauge flux, the shortest distance $m^{(x)}$ and $m^{(y)}$ can hop by in their longitudinal direction is by one lattice spacing. On the other hand, the mobility of $m^{(x)}$ and $m^{(y)}$ in their transverse directions is determined by Eq.~\eqref{eqn:mxNconstraintePart} and~\eqref{eqn:myNconstraintePart}, respectively. Only edges at a minimum $N$ lattice spaces away in their orthogonal direction are associated with the same gauge flux, and therefore the shortest distance $m^{(x)}$ and $m^{(y)}$ excitations can hop by in their transverse direction is $N$ lattice spaces.

As for the $e$ particles, their mobility is determined by only Eq.~\eqref{eqn:chargeNconstraintePart}. The closest distance two lattice site associated with the same gauge charge are is $N$ lattice spacing. Therefore the shortest distance $e$ particles can hop by in both the $x$ and $y$ direction is $N$ lattice spaces.

In addition to the elementary excitations $e$, $m^{(x)}$, and $m^{(y)}$, this analysis can be applied to composite excitations. Indeed, particular nontrivial excitations made of the $e$ particles and $\vec{m}$ particles carry a position-independent gauge charge and flux and can therefore move freely. For instance, consider the $m$ vector ``dipole''  corresponding to the lattice site $(x,y)$:
\begin{equation}\label{eqn:gExcitation}
    g_{x,y} = \bar{m}^{(x)}_{x,y}\otimes m_{x,y+1}^{(x)}.
\end{equation}
From Eq.~\eqref{eqn:mChargeLatDep}, as represented on the anyon lattice, it always carries the position-independent gauge flux ${\mathfrak{g}_{x,y} = \mathfrak{g}}$ and therefore its mobility is unrestricted. As for composite excitations made up of $e$ particles, from Eq.~\eqref{eqn:eChargeLatDep} the $e$ $x$-dipole
\begin{equation}\label{eqn:pxExcitation}
    p^{(x)}_{x,y} = \bar{e}_{x,y}\otimes e_{x+1,y}
\end{equation}
always carries the position-independent gauge charge ${\mathfrak{p}^{(x)}_{x,y} = \mathfrak{p}^{x}}$. Similarly, the $e$ $y$-dipole
\begin{equation}\label{eqn:pyExcitation}
    p^{(y)}_{x,y} = \bar{e}_{x,y}\otimes e_{x,y+1}
\end{equation}
always carries the position-independent gauge charge ${\mathfrak{p}^{(y)}_{x,y} = \mathfrak{p}^{y}}$. Therefore, both $p^{(x)}$ and $p^{(y)}$ are completely mobile.

The mobility of $e$, $m^{(x)}$, and $m^{(y)}$ excitations are closely connected to the conservation laws of the continuum $U(1)$ tensor gauge theory given by Eqs.~\eqref{eqn:dipCons} and~\eqref{eqn:rCrossBcons}. Indeed, these cause the $U(1)$ gauge charges (the $e$ particles before Higgsing) to be fractons and $U(1)$ magnetic flux loops (the $m^{(x)}$ and $m^{(y)}$ particles before Higgsing) to be lineons. The subdimensional behavior of the $U(1)$ case is also captured by the position-dependent gauge charge/flux anyon picture. The expressions for $\mathfrak{e}_{x,y}$, $\mathfrak{m}^{(x)}_{x,y}$, and $\mathfrak{m}^{(y)}_{x,y}$ given by Eqs.~\eqref{eqn:mChargeLatDep} and~\eqref{eqn:eChargeLatDep} apply independent to whether or not the gauge charges/fluxes are $\mathbb{Z}_{N}$ charges/fluxes as their derivation never used the fact that $N$-charges and $N$-fluxes condense. Then, for the $U(1)$ case the expression for $\mathfrak{e}_{x,y}$ is different for each lattice site, and therefore $U(1)$ $e$ particles cannot move (hence, they're fractons). As for $\mathfrak{m}^{(x)}_{x,y}$ and $\mathfrak{m}^{(y)}_{x,y}$ in the $U(1)$ case, because of Eqs.~\eqref{eqn:gaugeFluxCons1} and~\eqref{eqn:gaugeFluxCons2} they can still move by one lattice spacing in their longitudinal direction, but cannot move in their transverse direction (hence they're lineons). Upon Higgsing, since the $U(1)$ charges and fluxes become $\mathbb{Z}_{N}$ charges and fluxes, the equivalence relation~\eqref{eqn:chargeNconstraint} applies which allows a process where excitations hop by $N$ lattice sites in the direction they previously could not move.

While the position-dependent excitation picture recovers the mobility of $e$ and $\vec{m}$ excitations and their composite objects, the above analysis only concludes that there exists local string operators. However, the structure of these string operators is important in determining the low-energy dynamics. Indeed, when subdimensional particles condense in a Higgs phase, they can gain mobility only if all of the additional excited particles that usually prevent their movement are perfectly absorbed into the condensate~\cite{PhysRevB.96.195139, PhysRevB.97.235112, PhysRevB.104.165121, PhysRevB.98.035111, arxiv.2207.00409}. As such, the lattice string operators can be rather complicated. Indeed, the string operators that hops an $e$ particle from $(x,y)$ to $(x+N,y)$ or $(x,y+N)$, respectively, are~\cite{PhysRevB.97.235112, PhysRevB.98.035111, oh2021path}
\begin{equation}\label{eqn:eParticleStringOps}
\begin{aligned}
    W^{(e,x)}_{x,y} &= \prod_{i=0}^{N-1}\left(X^{\dagger}_{1,x+i,y}\right)^{i},\\
    W^{(e,y)}_{x,y} &= \prod_{i=0}^{N-1}\left(X^{\dagger}_{2,x,y+i}\right)^{i},
\end{aligned}
\end{equation}
and the string operators that hop $m^{(x)}$ and $m^{(y)}$ by $N$ lattice spaces in their transverse directions are
\begin{equation}
\begin{aligned}
    W^{(m^{(x)},y)}_{x,y} &= \prod_{i=0}^{N-1}Z_{3,x,y+i}\left(Z_{1,x,y+i}Z^{\dagger}_{1,x+1,y+i}\right)^{i},\\
    W^{(m^{(y)},x)}_{x,y} &= \prod_{i=0}^{N-1}Z_{3,x+i,y}\left(Z_{2,x+i,y}Z^{\dagger}_{2,x+i,y+1}\right)^{i}.
\end{aligned}
\end{equation}
On the other hand, the string operators that hop excitations in directions they're always mobile are much simpler. Indeed, the string operators to hop $m^{(x)}$ and $m^{(y)}$ in their longitudinal directions by one lattice site are simply
\begin{equation}
\begin{aligned}
    W^{(m^{(x)},x)}_{x,y} &= Z^{\dagger}_{2,x+1,y},\\
    W^{(m^{(y)},y)}_{x,y} &= Z^{\dagger}_{1,x,y+1}.
\end{aligned}
\end{equation}
Similarly, the string operators that hop the $e$ dipoles $p^{(x)}_{x,y}$ and $p^{(y)}_{x,y}$ by one lattice site are
\begin{equation}
\begin{aligned}
    W^{(p^{(x)},x)}_{x,y} &= X_{1,x,y},\\
    W^{(p^{(x)},y)}_{x,y} &= X_{3,x,y},\\
    W^{(p^{(y)},x)}_{x,y} &= X_{3,x,y},\\
    W^{(p^{(y)},y)}_{x,y} &= X_{2,x,y},
\end{aligned}
\end{equation}
and the ones for the $m$ vector dipole $g_{x,y}$ are
\begin{equation}
\begin{aligned}
    W^{(g,x)}_{x,y} &= Z_{2,x+1,y}Z^{\dagger}_{2,x+1,y+1},\\
    W^{(g,y)}_{x,y} &= Z_{3,x,y+1}Z_{1,x+1,y+1}Z^{\dagger}_{1,x,y+1}Z^{\dagger}_{3,x,y}.
\end{aligned}
\end{equation}

The rank-2 toric code Hamiltonian Eq.~\eqref{eqn:Hamiltonian} does not have dynamics, but by adding off-diagonal terms we can consider the corresponding low-energy effective Hamiltonian describing the induced dynamical processes. The leading order terms for small off-diagonal elements will consists of quantities made of the fewest operators. The minimum number of operators used to hop an $e$ particles by $N$ lattice sites is
\begin{equation}
    L_{e}(N) = 
    \begin{cases}
        N^{2}/4,\quad &N\text{ is even}\\
        (N^{2}-1)/4,\quad &N\text{ is odd}
    \end{cases},
\end{equation}
where as the minimum number of operators to hop $m^{(x)}$ and $m^{(y)}$ in their transverse directions, respectively, is
\begin{equation}
    L_{m}(N) = N + 2L_{e}(N).
\end{equation}
For $N>2$, the leading order dynamical processes are $\vec{m}$ particles moving in one direction and $p$ dipoles moving freely. In the low-energy effective Hamiltonian, the $e$ particles are therefore pseudo-fractons while the $\vec{m}$ particles are pseudo-lineons, and therefore there are pseudo-subdimensional (subdimensional at low-energies) particles in $2+1$d.

\subsubsection{Position-Dependent Braiding Statistics}\label{sec:braiding}

An interesting consequence of the $e$ and $\vec{m}$ particles carrying position dependent gauge charge and flux is that their mutual braiding statistics become position-dependent. While the position dependency of braiding-statistics can be inferred directly from the string operators~\cite{oh2021path}, the result seems rather magical. However, it becomes much more intuitive when understood as a consequence of anyons from different lattice sites carrying different gauge charge/flux. Furthermore, considering the excitations' braiding statistics will also be useful later on in Section~\ref{sec:CSthy} when we develop a mutual Chern-Simons theory as the low-energy effective field theory for the system.

The elementary excitations $e$, $m^{(x)}$, and $m^{(y)}$ are all bosons and therefore have trivial self-statistics. However, their nontrivial mutual statistics make them abelian anyons. Indeed, braiding $e_{x_{e},y_{e}}$ counterclockwise around either $m^{(x)}_{x_{m},y_{m}}$ or $m^{(y)}_{x_{m},y_{m}}$ will cause the many-body wave function to pick up a phase that depends on the initial coordinates $(x_{e},y_{e})$ and $(x_{m},y_{m})$. The phase accumulated from this can be found by first finding the braiding statistics between excitations carry the basis vectors of the anyon lattice. Then, using the expressions for $\mathfrak{e}_{x,y}$, $\mathfrak{m}^{(x)}_{x,y}$, and $\mathfrak{m}^{(y)}_{x,y}$ in Eqs.~\eqref{eqn:mChargeLatDep} and~\eqref{eqn:eChargeLatDep}, the braiding statistics between any elementary or composite excitations can be found.

First recall that as described by Eq.~\eqref{eq:basisgaugeflux}, the basis gauge fluxes carry the gauge flux of $m_{0,0}^{(x)}$, $m_{0,0}^{(y)}$, and $m_{0,1}^{(x)}$ and from Eq.~\eqref{eq:basisgaugecharge} the basis gauge charges carry the gauge charge of $e_{0,0}$, $e_{1,0}$, and $e_{0,1}$. Therefore, finding the mutual braiding statistics for the anyon lattice basis amounts to finding the mutual statistics between the $m_{0,0}^{(x)}$, $m_{0,0}^{(y)}$, and $m_{0,1}^{(x)}$ and $e_{0,0}$, $e_{1,0}$, and $e_{0,1}$. Using the string operators it is straight forward to find their mutual statistics from the relations ${Z_{j}X_{i} = \omega^{\delta_{i,j}}X_{i}Z_{j}}$ where ${\omega \equiv \exp[2\pi \mathrm{i} /N]}$.

Indeed, using the $e$ particle's string operators given by Eq.~\eqref{eqn:eParticleStringOps}, braiding $e_{0,0}$ around $m^{(x)}_{0,1}$, and $e_{1,0}$ around $m^{(x)}_{0,1}$ or $m^{(y)}_{0,0}$ all cause the many-body wave function to pick up the phase $\omega$ due to the relation ${X_{i}Z_{i}^{\dagger} = \omega Z_{i}^{\dagger}X_{i}}$. However, braiding $e_{0,1}$ around $m^{(x)}_{0,0}$ instead causes the many-body wave function to pick up the phase $\omega^{-1}$ due to the relation ${X_{i}^{\dagger}Z_{i}^{\dagger} = \omega^{-1}Z_{i}^{\dagger}X_{i}^{\dagger}}$. From these, we can find the braiding statistics between excitations carrying the basis gauge charges and fluxes, which is summarize in table~\ref{table:ExchangeStatistics}.

From the braiding statistics shown in table~\ref{table:ExchangeStatistics} and using Eqs.~\eqref{eqn:mChargeLatDep} and~\eqref{eqn:eChargeLatDep}, it is straight forward to find the mutual statistics between $e$ and $\vec{m}$ particles at any site. Indeed, braiding a single $e_{x_{e},y_{e}}$ particle around a single $m^{(x)}_{x_{m},y_{m}}$ counter clockwise, the accumulated phase is $\omega^{-(y_{e} - y_{m})}$. Therefore if there are $\ell_{e}$ units of $\mathfrak{e}_{x_{e},y_{e}}$ gauge charge and $\ell_{m^{(x)}}$ units of $\mathfrak{m}^{(x)}_{x_{m},y_{m}}$ gauge flux, the total phase is 
\begin{equation}
    \exp[\mathrm{i}\theta_{e,m^{(x)}}(x_{e},y_{e},x_{m},y_{m})] = \omega^{-\ell_{e}\ell_{m^{(x)}}(y_{e} - y_{m})}.
\end{equation}
Similarly, braiding a single $e_{x_{e},y_{e}}$ around a single $m^{(y)}_{x_{m},y_{m}}$, the accumulated phase is $\omega^{x_{e} - x_{m}}$. Therefore, given that there are $\ell_{e}$ units of $\mathfrak{e}_{x_{e},y_{e}}$ gauge charge and $\ell_{m^{(y)}}$ units of $\mathfrak{m}^{(y)}_{x_{m},y_{m}}$ gauge flux, the total phase becomes 
\begin{equation}
    \exp[\mathrm{i}\theta_{e,m^{(y)}}(x_{e},y_{e},x_{m},y_{m})] = \omega^{\ell_{e}\ell_{m^{(y)}}(x_{e} - x_{m})}.
\end{equation}
Because $\omega^{N} =1$, the position-dependent phase only depends on the excitations' relative positions in the $N \times N$ lattice unit cell. We note that from these expression, the braiding statistics of any composite excitations can also be readily found. Furthermore, these phases are in agreement with the results found in Refs.~\cite{oh2021path, oh2022effective}, validating our expressions for the position-dependent gauge charge and flux carried by the elementary excitations.

\begin{table}[t!]
\begin{ruledtabular}
{\renewcommand{\arraystretch}{1.5}
\begin{tabular}{c|ccc}
Braiding Statistics &  $\mathfrak{m}^{x}$ & $\mathfrak{m}^{y}$ & $\mathfrak{g}$ \\
\hline
$\mathfrak{e}$    & $1$   &   $1$     &     $\omega$         \\
$\mathfrak{p}^{x}$    & $1$   & $\omega$  &     $1$         \\
$\mathfrak{p}^{y}$    &    $\omega^{-1}$     & $1$  & $1$   \\
\end{tabular}}
\end{ruledtabular}
\caption{%
The $e_{x,y}$, $m^{(x)}_{x,y}$, and $m^{(y)}_{x,y}$ excitations of the rank-2 toric code have nontrivial mutual statistics and pick up phase factor from braiding $e_{x,y}$ around $m^{(x)}_{x,y}$ or $m^{(y)}_{x,y}$. The $e$ excitations carry basis gauge charges $\mathfrak{e}$ $\mathfrak{p}^{x}$, and $\mathfrak{p}^{y}$ and the $m^{(x)}$ and $m^{(x)}$ excitations carry basis gauge flux $\mathfrak{m}^{x}$, $\mathfrak{m}^{y}$, and $\mathfrak{g}$ (see Eqs.~\eqref{eqn:mChargeLatDep} and~\eqref{eqn:eChargeLatDep}). This table shows the phases picked up from braiding excitations carrying a single unit of each gauge charge/flux, with ${\omega \equiv \exp[2\pi \mathrm{i} /N]}$.
}
\label{table:ExchangeStatistics}
\end{table}

\subsubsection{Lattice Transformations Effect on the Anyon Lattice}\label{sec:latticeTransonAnyonLat}

So far, we have seen that from the fusion rules, the elementary excitations' vector representations in the anyon lattice depends on their position. This restricted their mobility and enriched them with position-dependent braiding statistics. Furthermore, because the anyon lattice is coupled to the direct lattice, this also means that lattice transformations induce transformations on the anyon lattice. We'll investigate these transformations in this section, further revealing the rich mixing between symmetry and the topological order in the rank-2 toric code.

The space group for the square lattice can be generated by a $4$-fold rotation, a mirror reflection, and translations in the $x$ and $y$ directions. Under the point group part of the space group, the excitations charge transform according to Eq.~\eqref{eqn:excitationsLatticeTrans}. While under the two translations, only their coordinates of the gauge charge/flux is changed (e.g., $T_{y}:~\mathfrak{e}_{x,y} \to \mathfrak{e}_{x,y+1}$ or $T_{x}:~\mathfrak{m}^{(y)}_{x,y} \to \mathfrak{m}^{(y)}_{x+1,y}$). Because lattice transformations change the position of excitations, they also induce a transformation on the gauge charge/flux it carries.

Consider the 4-fold rotation $C_{4}$ that rotates the lattice counterclockwise by $\pi/2$ about the lattice site $(0,0)$. According to Eq.\eqref{eqn:c4transLatticeCharge}, $C_{4}$ transforms the gauge flux $m^{(x)}$ and $m^{(y)}$ excitations carry as
\begin{align*}
    C_{4}:\quad\mathfrak{m}^{(x)}_{x,y}&\to \mathfrak{m}_{-y,x}^{(y)} = \mathfrak{m}^{y} + y~ \mathfrak{g},\\
    C_{4}:\quad\mathfrak{m}^{(y)}_{x,y}&\to -\mathfrak{m}^{(x)}_{-y-1,x} =-\mathfrak{m}^{x} - x\mathfrak{g}.
\end{align*}
However, considering Eq.~\eqref{eqn:mChargeLatDep}, this is equivalent to starting with $\mathfrak{m}^{(x)}_{x,y}$ and $\mathfrak{m}^{(y)}_{x,y}$ and instead of transforming $(x,y)$, transforming the basis gauge fluxes as
\begin{equation}\label{eqn:c4gaugeFluxBasis}
    \begin{aligned}
    \mathfrak{m}^{x} &\to \mathfrak{m}^{y},\\
    C_{4}:\quad \mathfrak{m}^{y} &\to -\mathfrak{m}^{x},\\
    \mathfrak{g} &\to \mathfrak{g}.
\end{aligned}
\end{equation}
As for the $e$ particles, under the rotation $C_{4}$ the gauge charge carried by $e_{x,y}$ transforms as
\begin{equation*}
    C_{4}:\quad\mathfrak{e}_{x,y} \to \mathfrak{e}_{-y,x} =\mathfrak{e} -y~\mathfrak{p}^{x} + x~\mathfrak{p}^{y}.
\end{equation*}
Comparison to Eq.~\eqref{eqn:eChargeLatDep}, the transformation is equivalent to starting with $\mathfrak{e}_{x,y}$ and transforming the gauge charge basis as
\begin{equation}\label{eqn:c4gaugeChargeBasis}
    \begin{aligned}
    \mathfrak{e} &\to \mathfrak{e},\\
    C_{4}:\quad\mathfrak{p}^{x} &\to \mathfrak{p}^{y},\\
    \mathfrak{p}^{y} &\to -\mathfrak{p}^{x}.
\end{aligned}
\end{equation}
Therefore, a rotation of the lattice indeed induces a transformation on the anyon lattice.

Consider a generic excitation represented by the anyon lattice vector $\vec{\ell}$, given by Eq.~\eqref{eqn:genericExcitationOnAnyonLattice}. Then, from Eq.~\eqref{eqn:c4gaugeFluxBasis} and Eq.~\eqref{eqn:c4gaugeChargeBasis}, the lattice rotation $C_{4}$ induces a transformation on $\vec{\ell}$ represented as
\begin{equation*}
    C_{4}^{(\mathcal{A})}:\vec{\ell}
    \to
    \begin{pmatrix}
         -\ell_{2}~\operatorname{mod}~N \\
         \ell_{1} \\
         \ell_{3}
    \end{pmatrix}
    \oplus
    \begin{pmatrix}
         \ell_{4} \\
         -\ell_{6}~\operatorname{mod}~N \\
         \ell_{5}
    \end{pmatrix}
\end{equation*}
The anyon lattice vector's components are $\ell_{i}\in\mathbb{Z}_{N}$ and therefore in order for the transformed $\ell$ to remain in $\mathbb{Z}^{6}_{N}$, the coefficients picking up a minus sign have to be mod $N$. This makes the transformation on the anyon lattice vector nonlinear. Nevertheless, we note that because any integers $a$ and $b$ satisfy ${-(-a~\operatorname{mod}~b)~\operatorname{mod}~b = a~\operatorname{mod}~b}$, this correctly satisfies $(C_{4}^{(\mathcal{A})})^{4} = 1$

The calculation and reasoning can be repeated for the other three
transformations that generate the space group of the square lattice. Indeed,
for a mirror reflection about the horizontal line $y = 0$, the induced
transformation on the anyon lattice is represented by
\begin{equation*}
    M_{x}^{(\mathcal{A})}:\vec{\ell}
    \to
    \begin{pmatrix}
         \ell_{1} \\
         -\ell_{2}~\operatorname{mod}~N \\
         -\ell_{3}~\operatorname{mod}~N
    \end{pmatrix}
    \oplus
    \begin{pmatrix}
         \ell_{4} \\
         \ell_{5} \\
         -\ell_{6}~\operatorname{mod}~N
    \end{pmatrix}
\end{equation*}
Note that this correctly satisfies $(M_{x}^{(\mathcal{A})})^{2} = 1$. 
As for translations in the $x$ and $y$ direction by one lattice spacing, the
matrices acting on $\vec{\ell}$ are represented by
\begin{equation}\label{eq:txyanyonlatticetrans}
\begin{aligned}
    T_{x}^{(\mathcal{A})}:\vec{\ell}
    &\to
    \begin{pmatrix}
         \ell_{1} \\
         \ell_{2} \\
         \ell_{3}-\ell_{2}~\operatorname{mod}~N
    \end{pmatrix}
    \oplus
    \begin{pmatrix}
         \ell_{4} \\
         \ell_{4}+\ell_{5}~\operatorname{mod}~N \\
         \ell_{6}
    \end{pmatrix},\\[0.5em]
    T_{y}^{(\mathcal{A})}:\vec{\ell}
    &\to
    \begin{pmatrix}
         \ell_{1} \\
         \ell_{2} \\
         \ell_{1}+\ell_{3}~\operatorname{mod}~N
    \end{pmatrix}
    \oplus
    \begin{pmatrix}
         \ell_{4} \\
         \ell_{5} \\
         \ell_{4}+\ell_{6}~\operatorname{mod}~N
    \end{pmatrix}.
\end{aligned}    
\end{equation}
As a consequence of Eq.~\eqref{eqn:chargeNconstraint}, both $T_{x}^{(\mathcal{A})}$ and $T_{y}^{(\mathcal{A})}$ transformations satisfy ${(T_{x}^{(\mathcal{A})})^{N} = (T_{y}^{(\mathcal{A})})^{N} = 1}$, which can easily be confirmed from the above expression.

All of the symmetry elements that generate the square lattice space group act on the anyon lattice vectors as non-linear transformations. However, for odd $N$, if we choose the range of $\ell_i$ to be ${-\frac{N-1}{2},-\frac{N-1}{2}+1,\cdots,0,\cdots,\frac{N-1}{2}}$, then we can drop $\operatorname{mod}~N$ in the above transformations of $C_4^{(\mathcal{A})}$ and $M_{x}^{(\mathcal{A})}$. In this case, $C_4^{(\mathcal{A})}$ and $M_{x}^{(\mathcal{A})}$ are linear transformations in the 6-dimensional anyon lattice. This implies that the lattice space group transformations $C_4$ and $M_{x}$ can be realized as a linear transformations on the six $U(1)$ gauge fields in the effective Chern-Simons theory discussed in Section~\ref{sec:CSthy}.

But for the lattice translations, we cannot find a range of $\ell_i$'s to linearize $T_x^{(\mathcal{A})}$ and $T_y^{(\mathcal{A})}$.  This implies that that the lattice translations cannot be realized as transformations on the six $U(1)$ gauge fields in the effective Chern-Simons theory in Section~\ref{sec:CSthy}. On the other hand, if the $\operatorname{mod}~N$ equivalence of the $U(1)$ gauge charges could be implemented in the  effective Chern-Simons theory, then the lattice translations could be realized as transformations on the $U(1)$ gauge fields.

One may try to linearize the space group transformation on anyon types by enlarging the number of basis anyon lattice vectors. If the anyon lattice vector space is spanned by the number of indistinguishable elementary excitations, so there are $N^{2}$ gauge charge anyon lattice basis vectors and $2N$ gauge flux basis vectors, the lattice transformations would act on the anyon lattice vectors linearly. However, this is not a legal basis because, as we saw in Section~\ref{sec:anyonLattice}, these anyon lattice vectors are linearly dependent.

As discussed in Section~\ref{sec:anyonLattice}, the rank-2 toric code has $\mathbb{Z}^{3}_{N}$ topological order that is enriched by the square lattice's space group. In Section~\ref{sec:intro}, we reviewed how there are unconventional SET orders in which symmetry elements additionally exchange inequivalent anyon types. The transformations on the anyon lattice vectors above describe these nontrivial automorphisms.

It is interesting to note that the fact symmetry elements exchange inequivalent anyon types is not explicit from the rank-2 toric code Hamiltonian. For instance, Wen's plaquette model~\cite{PhysRevB.78.155134, PhysRevB.86.161107}, $\mathbb{Z}_{2}$-charges live on $A$ plaquettes while $\mathbb{Z}_{2}$-fluxes live on $B$ plaquettes, is in an unconventional SET phase. From the Hamiltonian, it is explicit that any lattice transformations that exchange $A$ and $B$ plaquettes (e.b., translations on by lattice spacing) will also exchange these anyon types. Here, lattice symmetries exchange inequivalent flavors of $\mathbb{Z}_{N}$-charges or inequivalent flavors of $\mathbb{Z}_{N}$-fluxes, which are, respectively, excited by the same local operators. Their inequivalencies arise from their nontrivial fusion rules, which are influenced by the conservation laws in the continuum tensor gauge theory: the fusion rules for exciting $e$ particles are the simplest ones that conserve dipole moment, Eq.~\eqref{eqn:dipCons}, and the fusion rules for exciting $\vec{m}$ particles are the simplest ones that conserve ``magnetic angular momentum,'' Eq.~\eqref{eqn:rCrossBcons}.

\subsection{Global Anyon Equivalence Relations}\label{sec:globalEquiv}

So far, we have considered effects arising from local operators (from the fusion rules studied) that were independent of the system's boundary conditions. In this section, we now consider the system with periodic boundary conditions. In particular, how the presence non-local operators wrapped around the nontrivial cycles of the torus affect the excitations in the rank-2 toric code. These global operators will give rise to new equivalence relations between anyon types. Indeed, for an $L_{x}\times L_{y}$ square lattice, periodic boundary conditions require that gauge charges and fluxes satisfy the global equivalence relations
\begin{equation}\label{eqn:PBC}
    \begin{aligned}
    \mathfrak{e}_{x,y} &\pbceq \mathfrak{e}_{x+L_{x},y} \pbceq \mathfrak{e}_{x,y + L_{y}},\\
    \vec{\mathfrak{m}}_{x,y} &\pbceq \vec{\mathfrak{m}}_{x + L_{x},y} \pbceq \vec{\mathfrak{m}}_{x,y + L_{y}}.
    \end{aligned}
\end{equation}
We use the notation $\pbceq$ and terminology ``global'' to denote that the equivalence relation is satisfied only given periodic boundary conditions and affects only non-local operators that wind around the system. Using Eqs.~\eqref{eqn:mChargeLatDep} and~\eqref{eqn:eChargeLatDep}, these yield
\begin{equation}\label{eqn:pbcGlobalConstraints}
\begin{aligned}
    L_{x}~\mathfrak{p}^{x} &\pbceq 0,\hspace{40pt}
    L_{y}~\mathfrak{p}^{y} \pbceq 0,\\
    L_{x}~\mathfrak{g} &\pbceq 0,\hspace{40pt}L_{y}~\mathfrak{g} \pbceq 0.
\end{aligned}
\end{equation}

Both local and global operators can condense $N$ elementary excitations into the vacuum, arising from the property that ${X_{i}^{N} = Z_{i}^{N} = 1}$. This too, of course, applies for the nontrivial excitations $p^{(x)}$, $p^{(y)}$, and $g$. Eq.~\eqref{eqn:pbcGlobalConstraints} reveals that there exists operators that wind around the system in the $x$ direction that can cause $L_{x}$ of the $p^{(x)}$ and $L_{x}$ of the $g$ excitations to condense. And similarly, that there exists operators that wind around the system in the $y$ direction that can cause $L_{y}$ of the $p^{(y)}$ and $L_{y}$ of the $g$ excitations to condense. We will only be concerned with the exact form of these and similar non-local operators when it is required, and instead focus on the general consequences of their existence using the equivalence relations.

The above discussion implies that global operators can potentially condense fewer than $N$ charges into the vacuum. Indeed, the number of charges can be found by a repeated condensing algorithm. For instance, consider the simultaneous equivalence relations for $p^{(x)}$, which we'll call level zero of the procedure:
\begin{equation*}
    \text{Level $0$:}\hspace{20pt} N~\mathfrak{p}^{x} = 0,\hspace{15pt}L_{x}~\mathfrak{p}^{x} \pbceq 0.
\end{equation*}
Assuming $L_{x}>N$, consider $L_{x}$ of the $\mathfrak{p}^{x}$ charges. While according to Eq.~\eqref{eqn:pbcGlobalConstraints} this can be condensed into the vacuum, let's use that fact that $N~\mathfrak{p}^{x} = 0$ to instead condense only $N$ of them. Because $L_{x}$ of the $p^{(x)}$ anyons were equivalent to the trivial excitation, $(L_{x} - N)$ of them must also be equivalent to the trivial excitation and satisfy ${(L_{x}-N)~\mathfrak{p}^{x} \pbceq 0}$. Therefore, the simultaneous equivalence relations of level zero are updated to level one:
\begin{equation*}
    \text{Level $1$:}\hspace{20pt} N~\mathfrak{p}^{x} = 0, \hspace{15pt} (L_{x}-N)~\mathfrak{p}^{x} \pbceq 0.
\end{equation*}
If ${(L_{x}-N) >N}$ we'll progress to level 2a of the procedure by condensing $N$ more of the $\mathfrak{p}^{x}$ charges, leading to
\begin{equation*}
    \text{Level 2a:}\hspace{20pt} N~\mathfrak{p}^{x} = 0,\hspace{15pt}(L_{x}-2N)~\mathfrak{p}^{x} \pbceq 0.
\end{equation*}
On the other hand, if ${(L_{x}-N) < N}$, we instead start from ${N~\mathfrak{p}^{x} = 0}$ and condense ${(L_{x}-N)}$ of the $\mathfrak{p}^{x}$ charges. This gives the other possibility for level two of the procedure
\begin{equation*}
    \text{Level 2b:}\hspace{20pt} (2N - L_{x})~\mathfrak{p}^{x} \pbceq 0,\hspace{15pt}(L_{x}-N)~\mathfrak{p}^{x} \pbceq 0.
\end{equation*}
This repeated condensation procedure continues by subtracting the smaller integer from the larger integer of the two equivalence relations until they yield the same equivalence relation. Indeed, at this final level of the procedure
\begin{equation*}
    \text{Final Level:}\hspace{20pt} N_{\text{eff}}~\mathfrak{p}^{x} \pbceq 0,\hspace{15pt}N_{\text{eff}}~\mathfrak{p}^{x} \pbceq 0.
\end{equation*}
However, this is exactly Euclid's algorithm for finding the greatest common divisor ($\gcd$) between two integers~\cite{jones2012elementary}. Therefore, given $L_{x}$ and $N$ in level zero, the final level will always have $N_{\text{eff}} = \gcd(L_{x},N)$, and so the two simultaneous equivalence relations ${N~\mathfrak{p}^{x} = 0}$ and ${L_{x}~\mathfrak{p}^{x} \pbceq 0}$ imply the single one ${\gcd(L_{x},N)~\mathfrak{p}^{x} \pbceq 0}$.

The same condensing procedure can be repeated for the $\mathfrak{p}^{y}$ charge and the $\mathfrak{g}$ flux.  The only different is that for the $\mathfrak{g}$ flux, there are now three constraints that need to be simultaneously considered:
${N~\mathfrak{g} = 0}$, ${L_{x}~\mathfrak{g} \pbceq 0}$, and ${L_{y}~\mathfrak{g} \pbceq 0}$, and therefore the repeated condensation procedure will instead give ${\gcd(L_{x},L_{y},N)}$. Therefore, by taking into account that the elementary excitations are $\mathbb{Z}_{N}$ charges and fluxes, Eq.~\eqref{eqn:pbcGlobalConstraints} simplifies to
\begin{subequations}\label{eqn:pbcGlobalConstraints2}
\begin{align}
    \gcd(L_{x},N)~\mathfrak{p}^{x} &\pbceq 0,\\
    \gcd(L_{y},N)~\mathfrak{p}^{y} &\pbceq 0,\\
    \gcd(L_{x},L_{y},N)~\mathfrak{g} &\pbceq 0,
\end{align}
\end{subequations}

Local operators do not have access to the equivalence relations of Eq.~\eqref{eqn:pbcGlobalConstraints2} and therefore still view $\mathfrak{p}^{x}$ and $\mathfrak{p}^{y}$ as $\mathbb{Z}_{N}$-charges and $\mathfrak{g}$ as an $\mathbb{Z}_{N}$-flux. So, the anyon lattice $\mathcal{A}$ to local operators is still spanned by three $\mathbb{Z}_{N}$-charges and three $\mathbb{Z}_{N}$-fluxes. However, non-local operators utilizing the periodic boundary conditions can induces processes that condense $\gcd(L_{x},N)$ of the $\mathfrak{p}^{x}$ charges, $\gcd(L_{y},N)$ of the $\mathfrak{p}^{y}$ charges, and $\gcd(L_{x},L_{y},N)$ of the $\mathfrak{g}$ charges. Therefore, these global operators perceive the anyon lattice instead as
\begin{equation}\label{eqn:globalAnyonLattice}
    \mathcal{A} \pbceq \mathbb{Z}^{3}_{N}\otimes \mathbb{Z}_{\gcd(L_{x},N)}\otimes \mathbb{Z}_{\gcd(L_{y},N)}\otimes \mathbb{Z}_{\gcd(L_{x},L_{y},N)}. 
\end{equation}
The subgroup of the anyon lattice corresponding to only gauge charges according to non-local operators is
\begin{equation*}
    \mathcal{A}_{e} \pbceq \mathbb{Z}_{N}\otimes \mathbb{Z}_{\gcd(L_{x},N)}\otimes \mathbb{Z}_{\gcd(L_{y},N)},
\end{equation*}
while the subgroup of only gauge fluxes according to non-local operators is 
\begin{equation*}
    \mathcal{A}_{m} \pbceq \mathbb{Z}^{2}_{N}\otimes \mathbb{Z}_{\gcd(L_{x},L_{y},N)}. 
\end{equation*}
In this remainder of this section, we will consider two particularly interesting consequences of this result.

\subsubsection{Non-Local String Operators}

In Section~\ref{sec:mobility}, we found that as a consequence of the equivalence relations in Eq.~\eqref{eqn:chargeNconstraint}, there exists local operators that hop $e$ particles by $N$ lattice spaces in the $x$ and $y$ directions and that hop $m^{(x)}$ and $m^{(y)}$ particles by $N$ lattice spaces in their transverse directions. Similarly, as a result of the global equivalence relations in Eq.~\eqref{eqn:pbcGlobalConstraints2}, using Eqs.~\eqref{eqn:mChargeLatDep} and~\eqref{eqn:eChargeLatDep} implies the equivalence relations
\begin{subequations}\label{eqn:PBChopping}
    \begin{align}
        \mathfrak{e}_{x,y} &\pbceq \mathfrak{e}_{x+\gcd(L_{x},N),y} \pbceq \mathfrak{e}_{x,y + \gcd(L_{y},N)},\label{eqn:eParticlePBChopping}\\
        \mathfrak{m}^{(x)}_{x,y} &\pbceq \mathfrak{m}^{(x)}_{x,y + \gcd(L_{x},L_{y},N)},\\
        \mathfrak{m}^{(y)}_{x,y} &\pbceq \mathfrak{m}^{(y)}_{x + \gcd(L_{x},L_{y},N),y}.
    \end{align}
\end{subequations}
Therefore there exists non-local operators that hop $e$
particles in the $x$ direction by $\gcd(L_{x},N)$ sites and in the $y$ direction by $\gcd(L_{y},N)$, and that hop $m^{(x)}$ and $m^{(y)}$ particles $\gcd(L_{x},L_{y},N)$ lattice sites in their transverse directions, respectively\footnote{In the presence of periodic boundary conditions, the braiding statistics between a gauge charge and gauge flux can be come quite complicated. By utilizing the periodic boundary conditions, particles can braiding and pick up different phases than those described by the expressions in Section~\ref{sec:braiding}.}. Physically, these involve complicated processes where the excitations using local operators and winding around the system such that their net displacement is as described above.

For an $e$ particle, when $L_{i}$ ($i=x$ or $y$) is a multiple of $N$, hopping $e$ around the system once in the $i$-direction and returns to its position. Thus, as verified by Eq.~\eqref{eqn:eParticlePBChopping}, in this case global operators can still only hop $e$ by a net $N$ lattice sites. However, When $L_{x}$ or $L_{y}$ is not a multiple of $N$, after winding around the system once in that direction, the $e$ particle does not return to where it started. Instead, it arrives at a different lattice site within the $N\times N$ unit cell $\gcd(L_{i},N)$ sites away. Hence the non-local string operator that hops $e$ by $\gcd(L_{i},N)$ lattice sites is just the local string operator wrapped around the system once.

\begin{figure}[t!]
\centering
    \includegraphics[width=.48\textwidth]{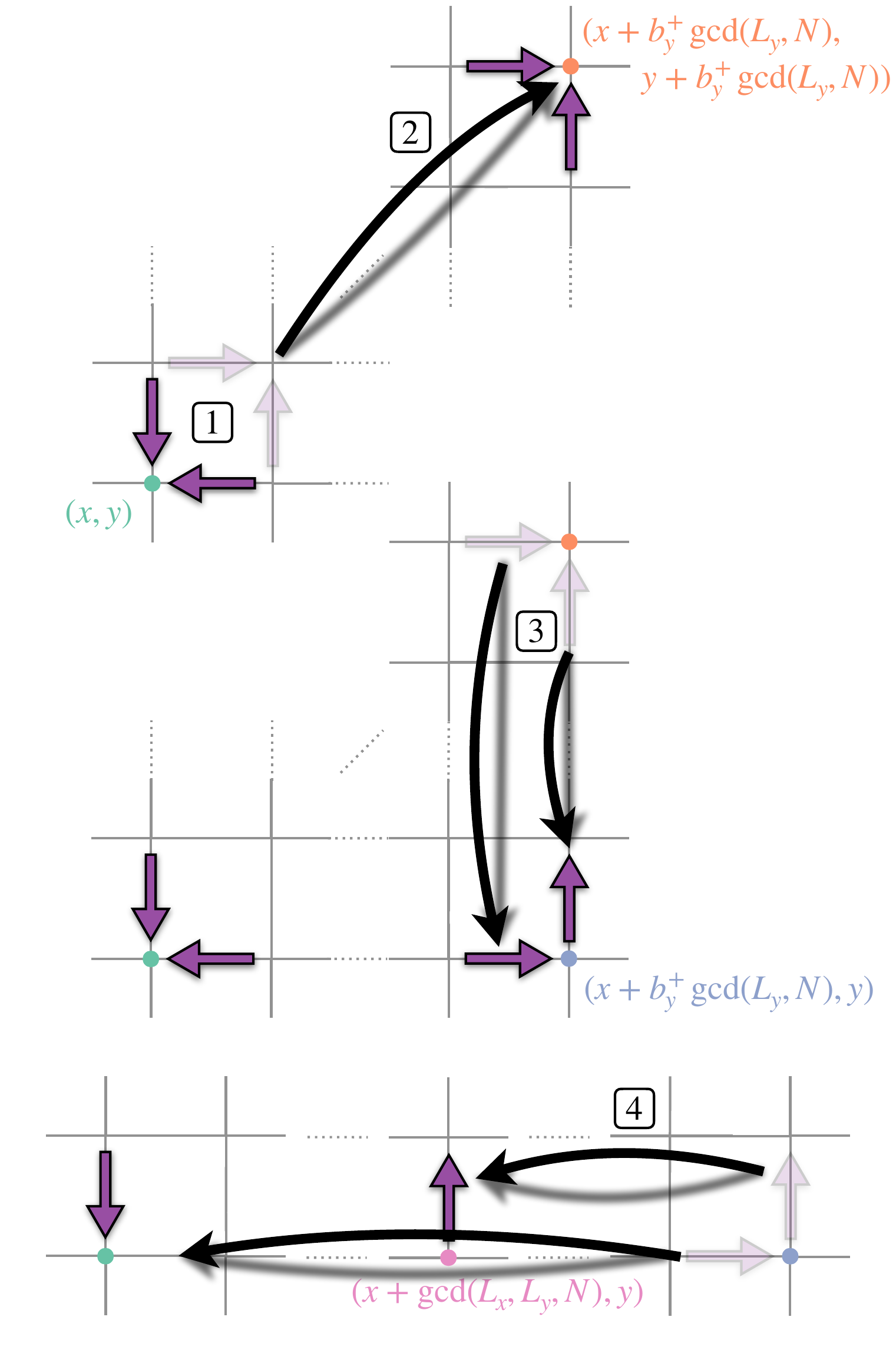}
    \caption{In the presence of periodic boundary conditions, $m^{x}$ and $m^{y}$ excitations can hop in the $y$ and $x$ directions, respectively, by $\gcd(L_{x},L_{y},N)$ lattice sites. This string operator is complicated, but can be constructed using a four-step recipe, as described in the main text. The top panel shows the first two steps, the middle panel shows the third step, and the bottom panel shows the fourth step of constructing this string operator for $m^{(y)}$. Relevant lattice sites are labeled once but color coded throughout the panels to avoid cumbersome labeling, and $b_{y}^{+}$ is a Bézout coefficient satisfying Eq.~\eqref{eqn:bezoutEq}
    }
    \label{fig:stringOp}
\end{figure}

The non-local string operators that hop an $m^{(x)}$ or $m^{(y)}$ particle by $\gcd(L_{x},L_{y},N)$ in their transverse directions are much more complicated. It is constructed using a four-step sequence of operators that excite $\vec{m}$ particles from vacuum and hop them throughout the system until the final operator is the desired non-string operator for hopping by $\gcd(L_{x},L_{y},N)$ sites. The sequence uses a result from elementary number theory known as Bézout's identity, which states that the $\gcd$ of two integers $a_{1}$ and $a_{2}$ can be written as~\cite{jones2012elementary}
\begin{equation*}
    \gcd(a_{1},a_{2}) = b_{1}a_{1} + b_{2}a_{2}.
\end{equation*}
There are an infinite number of ``Bézout coefficients'' $b_{1}$ and $b_{2}$, but one is always a non-negative integer and the other a non-positive integer\footnote{Given a pair of Bézout coefficients ($b_{1}$,$b_{2}$) for the integers $a_{1}$ and $a_{2}$, all other pairs are given by $\left(b_{1}-n\frac{a_{2}}{\gcd(a_{1},a_{2})}, ~b_{2}+n\frac{a_{1}}{\gcd(a_{1},a_{2})}\right)$, where $n\in\mathbb{Z}$.}. Using that
\begin{equation*}
    \gcd(L_{x},L_{y},N) = \gcd(~\gcd(L_{x},N),~\gcd(L_{y},N)~)
\end{equation*}
and applying Bézout's identity gives the decomposition
\begin{equation}\label{eqn:bezoutEq}
    \gcd(L_{x},L_{y},N) = b_{x}\gcd(L_{x},N) + b_{y}\gcd(L_{y},N).
\end{equation}
In constructing the string operator, we'll use two the minimal pairs of Bézout coefficients $(b^{\pm}_{x}, b^{\mp}_{y})$, which are defined as the smallest positive ($b^{+}_{x,y}$) and negative ($b^{-}_{x,y}$) Bézout coefficients. The recipe for the string operator that hops $m^{(y)}$ by $\gcd(L_{x},L_{y},N)$ sites in the $x$ direction is as follows (see Fig.~\ref{fig:stringOp} for a visualization of each step in the produce).

(1) First, using the fusion rule Eq.~\eqref{eqn:mFusionRules3}, excite the trivial excitation
    \begin{equation*}
        \bar{m}^{(x)}_{x,y}\otimes \bar{m}^{(y)}_{x,y} \otimes m^{(x)}_{x,y+1}\otimes m^{(y)}_{x+1,y}
    \end{equation*}
from the vacuum. The next three steps will involve hopping these four excitations throughout and around the system. Regardless of its position, we'll denote the excitation $\bar{m}^{(x)}_{x,y}$ as simply $\bar{m}^{(x)}$, $m^{(y)}_{x+1,y}$ as $m^{(y)}$, etc, without any risk for ambiguity.

(2) The composite object ${m^{(x)}_{x,y+1}\otimes m^{(y)}_{x+1,y}}$ can move freely along the ${\hat{x}+\hat{y}}$ direction. Indeed, from Eq.~\eqref{eqn:mChargeLatDep}, it carries gauge flux
\begin{equation*}
    \mathfrak{m}^{(x)}_{x,y+1} + \mathfrak{m}^{(y)}_{x+1,y} = \mathfrak{m}^{x} + \mathfrak{m}^{y} + (y-x)~\mathfrak{g}
\end{equation*}
which in invariant under translations in the ${\hat{x}+\hat{y}}$ direction. The second step is to hop ${m^{(x)}_{x,y+1}\otimes m^{(y)}_{x+1,y}}$ in the ${\hat{x}+\hat{y}}$ direction ${(b_{y}^{+}\gcd(L_{y},N)-1)}$ times.

(3) The excitation $m^{(y)}$ can always hop by one lattice site in the $y$ direction, and using a non-local operator $m^{(x)}$ can hop in the $y$ direction by $\gcd(L_{y},N)$ lattice spaces. The latter is for the same reason that with periodic boundary conditions an $e$ particle can hop by $\gcd(L_{i},N)$ in the $i$ direction. The third step is to hop $m^{(y)}$ in the $-\hat{y}$ direction ${(b_{y}^{+}\gcd(L_{y},N)-1)}$ times, and to hop $m^{(x)}$ by $\gcd(L_{y},N)$ lattice sites $b_{y}^{+}$ times in the $-\hat{y}$ direction.

(4) The fourth, and final, step is to first hop $m^{(x)}$ by one lattice site ${(b_{y}^{+}\gcd(L_{y},N)-1)}$ times in the $-\hat{x}$ direction such that it stops at site $(x,y)$ and annihilates with $\bar{m}^{(x)}$. Then hop $m^{(y)}$ by $\gcd(L_{x},N)$ lattice sites in the $-\hat{x}$ direction $(-b_{x}^{-})$ times. Because $b^{-}_{x}$ and $b^{+}_{y}$ are the Bézout coefficients of the decomposition Eq.~\eqref{eqn:bezoutEq}, $m^{(y)}$ will stop at position ${(x+\gcd(L_{x},L_{y},N),y)}$, a distance $\gcd(L_{x},L_{y},N)$ away from $\bar{m}^{(y)}$.

This recipe creates a string operator that hops $m^{(y)}$ from $(x,y)$ to $(x+\gcd(L_{x},L_{y},N),y)$. The string operator that hops $m^{(x)}$ from $(x,y)$ to $(x,y+\gcd(L_{x},L_{y},N))$ is created in a similar fashion. In the recipe, steps 1 and 2 involve only local operations that could have been done in the absence of periodic boundary conditions. It's important to note, however, that in order to hop $m^{(x)}$ by $\gcd(L_{y},N)$ lattice sites in the $y$ direction in step 3 (when $\gcd(L_{y},N)\neq N$) this string operator winds around the system. Similarly hopping $m^{(y)}$ by $\gcd{L_{x},N}$ lattice sites in the $x$ direction in step 4 (when $\gcd(L_{x},N)\neq N$) involves an operator that winds around the system. Indeed, hopping $m^{(y)}$ in the $x$ direction by $\gcd(L_{x},L_{y},N)$ lattice sites requires winding $m^{(y)}$ around the system $(-b_{x}^{-})$ times in the $x$ direction and winding a $m^{(x)}$ particle $b_{y}^{+}$ times around the system in the $y$ direction. Similarly, hopping $m^{(x)}$ in the $y$ direction by $\gcd(L_{x},L_{y},N)$ lattice sites requires winding $m^{(x)}$ around the system $(-b_{y}^{-})$ times in the $y$ direction and winding a $m^{(y)}$ particle $b_{x}^{+}$ times around the system in the $x$ direction.

\subsubsection{Ground State Degeneracy}\label{sec:anyonLatticeGSD}

Due to the global equivalence relations, non-local operators using periodic boundary conditions identify possibly fewer anyon types than local operators, depending on the system size. An effect of this is that the ground state degeneracy (GSD) becomes sensitive to the system size. In topologically ordered phases, degenerate ground states are distinguished by Wilson loop operators defined on the system's topologically nontrivial cycles and are subject to the low-energy constraint that defines the ground states subspace. So, because these Wilson loops distinguish possibly fewer anyon types, there can be fewer Wilson loop types and therefore the GSD can potentially be smaller than the naive guess.

\begin{figure}[t!]
\centering
    \includegraphics[width=.48\textwidth]{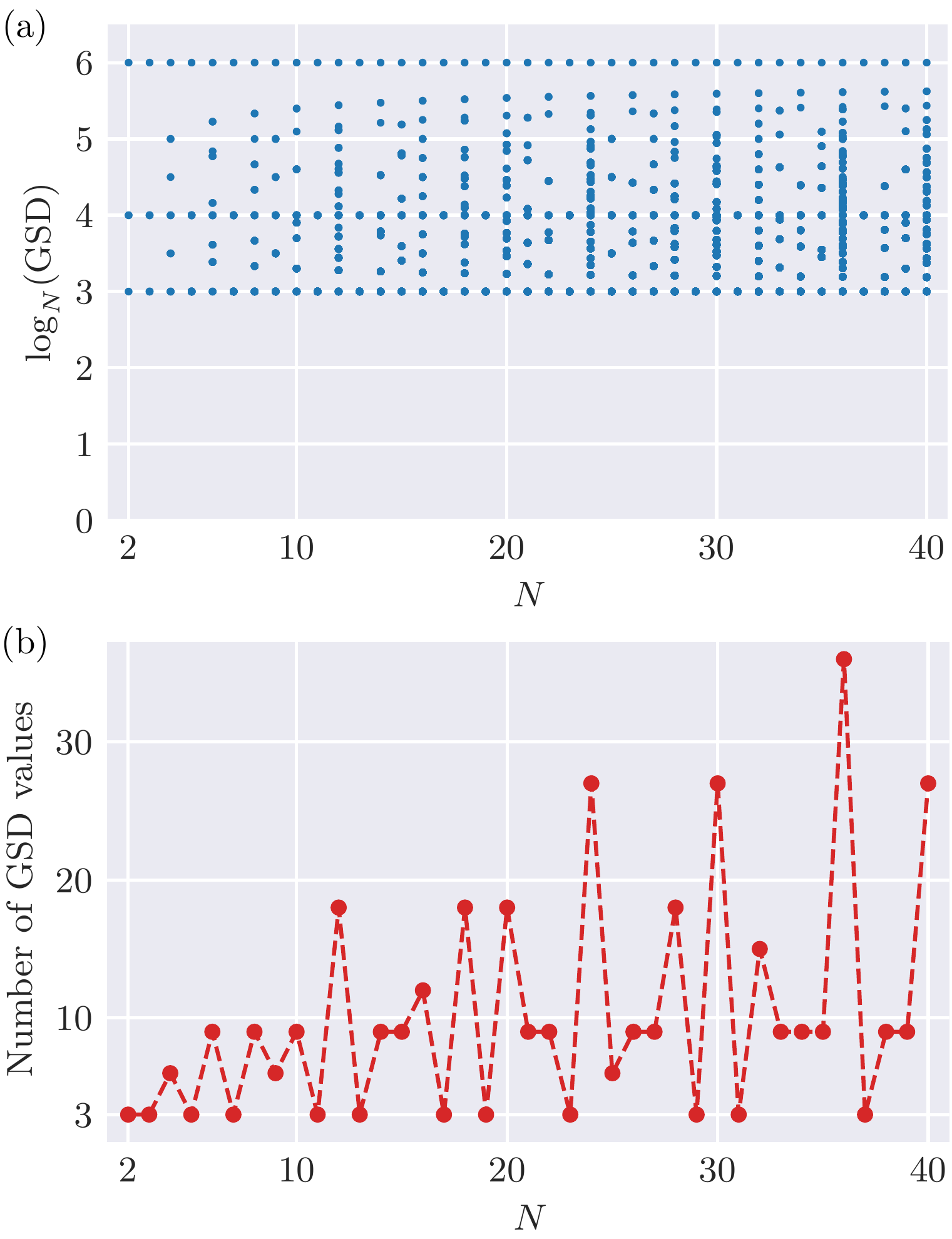}
    \caption{%
    The ground state degeneracy (GSD) of the rank-2 toric code is highly sensitive to the system's size (see Eq.~\eqref{eqn:GSD}).
    (a) For a fixed $N$, the GSD's value will change in the range $N^{3}\leq \text{GSD} \leq N^{6}$ as the system size $L_{x}\times L_{y}$ is changed.
    In the plot, for each value of $N$, the log base $N$ of all possible GSD values are plotted.
    (b) While there is always a finite number of different unique GSD values, the number them does not change monotonically with $N$ and is very chaotic. However, the number of GSD values is always at its minimum when $N$ is a prime, and the GSD only takes the values $N^{3}$, $N^{4}$, or $N^{6}$. In the plot, the dashed lines are drawn to guide the eye.
    }
    \label{fig:gsd}
\end{figure}

On a torus, the number of ground states is equal to the number of topological excitations --- the number of superselection sectors --- distinguishable by global Wilson loops~\cite{PhysRevB.90.115118, chan2016topological, PhysRevB.100.115147}.
The anyon lattice according to local operators is $\mathcal{A} = \mathbb{Z}^{6}_{N}$ and therefore local operators distinguish $N^{6}$ superselection sectors. However, according to global operators, the anyon lattice is given by Eq.~\eqref{eqn:globalAnyonLattice} and therefore global operators distinguish ${N^{3}\gcd(L_{x},N)\gcd(L_{y},N)\gcd(L_{x},L_{y},N)}$ superselection sectors. Because the Wilson loop operators perceive the latter number of sectors, the GSD is
\begin{equation}\label{eqn:GSD}
\text{GSD} = N^{3}\gcd(L_{x},N)\gcd(L_{y},N)\gcd(L_{x},L_{y},N).
\end{equation}
As shown in Fig.~\ref{fig:gsd}a, for fixed $N$ the GSD takes multiple different values in the range ${N^{3}\leq \text{GSD} \leq N^{6}}$ depending on the system size. Furthermore, the number of unique GSD values does not change monotonically with $N$, as shown in Fig.~\ref{fig:gsd}b. Instead, it depends on the number of divisors of $N$, which is always a minimum when $N$ is a prime number. Indeed, when $N$ is prime, $\gcd(L_{i},N)$ can either be $1$ or $N$, and therefore there are always only three different values of the GSD when $N$ is a prime: $N^{3}$, $N^{4}$, or $N^{6}$.

The extreme sensitivity of the GSD to the system's size prevents a well-defined continuum limit from existing. Indeed, changing the system size by a single lattice spacing, which is seemingly inconsequential after taking the continuum limit, leads to drastically different number of ground states. Therefore, any low-energy effective field theory must make explicit mention to the lattice spacing and number of lattice sites (which we'll indeed find to be true in Section~\ref{sec:effectiveAction}). This can be interpreted as a manifestation of UV/IR mixing~\cite{PhysRevB.103.195113, gorantla2022global}, where the low-energy (IR) physics cannot be decoupled from the high-energy (UV) physics. Here, low (high) energy referees to energies much smaller (larger) than the anyons' energy gaps. The UV/IR mixing arises because the anyon lattice and spatial lattice are coupled. The anyon lattice describes gapped excitations which are details of the UV theory. On the other hand, the system's size is only detectable by non-local (very long-wavelength) operators and is therefore an IR property. The global anyon equivalence relations relate the number of distinct anyon types (a UV property) to the system size (an IR property) and therefore a change in IR induces a change in the UV.

We emphasize that the GSD's sensitivity to the system size is a consequence of the global equivalent relations Eq.~\eqref{eqn:pbcGlobalConstraints}, which themselves were consequences of the fact that the anyons gauge charge and flux is position dependent. However, while not framed in the framework on lattice-dependent excitations, similar results have been seen in other lattice models. For instance, the GSD of Wen's $\mathbb{Z}_{2}$ plaquette model on a torus can either be $2$ or $4$, depending on if the system's linear size is even or odd~\cite{PhysRevB.78.155134, PhysRevB.86.161107}. Similarly, the GSD of the color code model~\cite{PhysRevLett.97.180501,PhysRevB.90.115118} is also sensitive to the system size of a hexagonal lattice based on whether or not the lattice is globally tricolorable. Furthermore, recently Seiberg \text{et al.} have found the GSD of several different models to depend on the the greatest common divisor between the system size and $N$~\cite{PhysRevB.103.195113, gorantla2022global}. In particular, as a consequence of possessing a $\mathbb{Z}_{N}$ global dipole symmetry, the ${1+1\text{d}}$ ``$\mathbb{Z}_{N}$ tensor gauge theory'' they study is closely related to the gauge charge sector of the rank-2 toric code. Indeed, ignoring the $y$-direction in our above analysis, the gauge-charge sector of the anyon lattice would be spanned by $\mathfrak{e}$ and $\mathfrak{p}^{x}$. Non-local operators perceive $\mathfrak{e}$ as a $\mathbb{Z}_{N}$-charge and $\mathfrak{p}^{x}$ as a $\mathbb{Z}_{\gcd(L_{x},N)}$-charge. Therefore, there are $N\gcd(L_{x},N)$ globally distinguishable anyons made out of only these gauge charges, which is the same number as the GSD found in the ${1+1\text{d}}$ ``$\mathbb{Z}_{N}$ tensor gauge theory'' studied in Ref.~\cite{gorantla2022global}. It would be interesting to see if the results from all of these mentioned models could also be understood in terms of position-dependent excitations.

\section{Low-Energy Effective Field Theory}

In Section~\ref{sec:bigR2TC}, we have found and studied the anyon lattice that describes the rank-2 toric code model. Using this vector space formalism, we have investigated the mobility of gapped excitations and their corresponding string operators, the position-dependency of their braiding statistics, and the ground state degeneracy on a torus. The ground state degeneracy, given by Eq.~\eqref{eqn:GSD} and plotted in Fig.~\ref{fig:gsd}, was extremely sensitive to the system's size. One may then wonder what, if any, long-wavelength effective field theory could have the rank-2 toric code as its UV regularization. In this section, we now develop such an effective theory that describes the topological order in the rank-2 toric code. The effective action we find reveals the UV/IR mixing that was hinted at in Section~\ref{sec:anyonLatticeGSD}: the IR theory's coupling constants explicitly depends on the number of unit cells in the UV regularized lattice.

\subsection{Mutual $U^{6}(1)$ Chern-Simons Theory}\label{sec:CSthy}

Mutual Chern-Simons theory is a powerful theoretical tool used to describe and characterize abelian topological orders in $2+1$d~\cite{PhysRevB.46.2290}. It acts as a long-wavelength effective field theory for energies below the excitations' gaps, described by the $2+1$d Minkowski spacetime action
\begin{equation}
S_{\text{MCS}} = \frac{K_{ij}}{4\pi}\int \mathrm{d}t~\mathrm{d}^{2}\bm{x}~ \epsilon^{\mu\nu\rho}a^{(i)}_{\mu}\partial_{\nu}a^{(j)}_{\rho}+\ldots,
\end{equation}
where $K$ is a symmetric integer matrix, $\epsilon^{\mu\nu\rho}$ is the antisymmetric Levi-Civita symbol, and the ellipsis denotes higher-order symmetry allowed terms, such as Maxwell terms. To connect this mutual Chern-Simons action with the 2+1d abelian topological order, we first introduce a compact $U(1)$ 3-vector gauge field ${a^{(i)} = (a^{(i)}_{0},\bm{a^{(i)}})}$ for each basis gauge charge and flux. We can then find the matrix elements of $K$ using the fact that the braiding statistics between two particles corresponding to the gauge fields $a^{(i)}$ and $a^{(j)}$ is given by $\theta_{ij} = 2\pi (K^{-1})_{ij}$.

For example, $\mathbb{Z}_{N}$ topological order is described by an effective $U^{2}(1)$ mutual Chern-Simons theory, where one gauge field corresponds to the $\mathbb{Z}_{N}$-charge and the other to the $\mathbb{Z}_{N}$-flux, and the $K$ matrix is given by ${K = N\sigma_{x}}$ (where $\sigma_{x}$ is the Pauli-$x$ matrix)~\cite{hansson2004superconductors, PhysRevB.78.155134}. The rank-2 toric code is instead described by a $U^{6}(1)$ mutual Chern-Simons theory. We'll consider the gauge fields (${a^{(1)},~a^{(2)},~a^{(3)}}$) corresponding to the three flavors of gauge flux (${\mathfrak{m}^{x},~\mathfrak{m}^{y},~\mathfrak{g}}$) and (${a^{(4)},~a^{(5)},~a^{(6)}}$) corresponding to the three independent flavors of $e$ particles (${\mathfrak{e},~\mathfrak{p}^{x},~\mathfrak{p}^{y}}$). Then, from the mutual statistics shown in table~\ref{table:ExchangeStatistics}, we can solve for the matrix elements of $K^{-1}$, and upon taking its inverse yields
\begin{equation}\label{eqn:kmatrix}
K = 
\begin{pmatrix}
0_{3} & C\\
C^{\top} & 0_{3}
\end{pmatrix}\quad\quad\quad
C = 
\begin{pmatrix}
0 & 0 & -N\\
0 & N & 0\\
N & 0 & 0
\end{pmatrix},
\end{equation}
where $0_{3}$ is a $3\times 3$ matrix of all zeros. Note that this correctly is an integer matrix\footnote{Here the matrix elements of $K$ are integers. When this is not the case, we note that the $K$ matrix can still be determined from the excitations' self and mutual-statistics, but auxiliary gauge fields must be included into the theory (see appendix C of Ref.~\cite{ma2020fractonic}).}. Furthermore, from the form of the $K$ matrix, this mutual Chern-Simons theory is equivalent to adding three mutual Chern-Simon terms describing $\mathbb{Z}_{N}$ topological order.
This agrees with the conclusion made in Section~\ref{sec:anyonLattice} and Refs.~\cite{PhysRevB.97.235112, PhysRevB.98.035111} that the rank-2 toric code has $\mathbb{Z}^{3}_{N}$ topological order.

However, the $K$ matrix alone does not fully characterize the topological order, the symmetry transformations and how they act on the gauge fields are also important. Indeed, for example, when translations act on the gauge fields such that they all satisfy periodic boundary conditions, the ground state degeneracy (GSD) on a torus is given by $|\det K|$~\cite{PhysRevB.46.2290, wesolowski1994multiple}. From the above $K$ matrix, this would give that the GSD is always $N^{6}$. However, in Section~\ref{sec:anyonLatticeGSD}, we found that this is only true when both $L_{x}$ and $L_{y}$ are multiples of $N$. When this is the case, the transformations acting on an anyon lattice vector induced by lattice translations, Eq.~\eqref{eq:txyanyonlatticetrans}, satisfy ${(T_{x}^{(\mathcal{A})})^{L_{x}}=1}$ and ${(T_{y}^{(\mathcal{A})})^{L_{y}}=1}$. Therefore, only when ${\gcd(L_{x},N) = \gcd(L_{y},N) = N}$ do the gauge fields all indeed satisfy periodic boundary conditions.

When this is not the case, this means that translations act on the gauge fields nontrivially, causing them to satisfy modified boundary conditions and therefore the GSD is no longer $|\det K|$. Indeed, for instance, when $L_{x}$ is not a multiple of $N$, $(T_{x}^{(\mathcal{A})})^{L_{x}}$ is no longer the identity and translations around the system induce an automorphism on the anyon lattice. Similar is true for translations around the system in the $y$ direction. This then causes the gauge fields to satisfy twisted periodic boundary conditions~\cite{PhysRevB.81.045323}, of the general form
\begin{equation}\label{eqn:TPBC}
    c^{x}_{i}a^{(i)}(x+L_{x},y)+c^{y}_{i}a^{(i)}(x,y+L_{y})+c_{i}a^{(i)}(x,y) = 0,
\end{equation}
where sum over $i$ is implied and $c^{x}_{i}$, $c^{y}_{i}$, and $c_{i}$ are integers.

To find these twisted boundary conditions, we need to know how translations act on the gauge fields. However, lattice transformations act on the anyon lattice vectors nonlinearly, which makes such a task nontrivial. Typically, for linear transformations, one finds how the gauge fields transform by considering a generic anyon lattice vector transforming under some linear transformation ${\vec{\ell}\to U\vec{\ell}}$. Assuming that $U$ corresponds to a symmetry of the theory, the gauge fields then transform as ${\vec{a}\to (U^{-1})^{\top}\vec{a}}$, where $(\vec{a})_{i} = a^{(i)}$. This is because in the field theory, a generic excitation is described by the term $(\vec{\ell}\cdot \vec{a}_{\mu} ) j^{\mu}$ in the effective Lagrangian density, and $\vec{a}$ transforms in such a way so $\vec{\ell}\cdot \vec{a}_{\mu}$ remains unchanged. Additionally, using this transformation of $\vec{a}$, in order for the mutual Chern-Simons term to remain unchanged by the transformation, we find the familiar result that the $K$ matrix must transform as $K\to UKU^{\top}$. Because the anyon-lattice transforms non-linearly, these familiar results using linear algebra no longer apply\footnote{It would be interesting if there exists a non-abelian Chern-Simons theory describing the same topological order but for which lattice transformations act on the gauge fields linearly.}.

While we cannot find how the lattice transformations act on the gauge fields of the Chern-Simons theory, we can still find an effective action using the mutual Chern-Simons theory with $K$ matrix~\eqref{eqn:kmatrix}. In what follows, we'll do so by considering the Holonomies of the torus in terms of these gauge fields, from which we find the zero modes of the gauge fields. Then, plugging in the gauge fields in terms of their zero modes into the mutual Chern-Simons theory yields an effective low-energy action.

\subsection{Holonomies of the Torus}\label{sec:holonomies}

The topological (global) contributions of the gauge fields are their zero modes, which are the holonomies of the torus. On a torus, for each gauge field there are two holonomies, so 
for the $U^{6}(1)$ mutual Chern-Simons theory written down in the previous section there are a total of twelve holonomies. These twelve holonomies act as a basis from which all other holonomies can be generated. We can find them by first considering gauge-invariant line-integrals along the nontrivial cycles of the torus. Because the gauge fields satisfy complicated twisted periodic boundary conditions, Eq.~\eqref{eqn:TPBC}, we'll see that in order to be gauge-invariant some of these line integrals will be integrating around the torus multiple times.

Physically, the holonomies are Wilson loops, corresponding to exciting particles and anti-particles and hopping the particles around the system until they return to where they started and can annihilate with the anti-particles back into the ground state. On the lattice, when excitations hop by $N$ lattice sites at a time they need to go around the system multiple times to return to the lattice site they started. In the field theory, having a particle return back to where it started is trivial since we're in the continuum limit. However, whether or not it can annihilate back into vacuum is subtle because of the twisted boundary conditions. One way to approach this is by considering how ${(T_{x}^{(\mathcal{A})})^{n_{x}L_{x}}}$ and ${(T_{y}^{(\mathcal{A})})^{n_{x} L_{y}}}$ act on the anyon lattice vectors for integers $n_{x}$ and $n_{y}$. While we do not know how $T_{x}^{(\mathcal{A})}$ and $T_{y}^{(\mathcal{A})}$ transform the gauge fields, we can instead find the smallest $n_{x}$ and $n_{y}$ such that ${(T_{x}^{(\mathcal{A})})^{n_{x}L_{x}}}$ and ${(T_{y}^{(\mathcal{A})})^{n_{x} L_{y}}}$ are the identity. If they act as the identity on anyon lattice vectors, this is the one case we do know how they act on the gauge fields: also as the identity. This amounts to ignoring the details of the twisted boundary conditions and instead finding out how many times the holonomy needs to go around the system in order to close (for the particles to annihilate back into vacuum).

Indeed, let's first consider the gauge charge sector. Recall that in Section~\ref{sec:mobility}, the non-trivial excitations $p^{(x)}$ and $p^{(y)}$, whose gauge charge corresponds to the gauge fields $a^{(5)}$ and $a^{(6)}$ respectively, can hop by one lattice site in every direction. In terms of the anyon lattice vectors, this means that the vector whose components are $\ell_{p_{x}}^{i} = \delta^{i,5}$ (for the $p^{x}$ excitation) or $\ell_{p_{y}}^{i} = \delta^{i,6}$ (for the $p^{y}$ excitation) are unchanged under lattice translations Eq.~\eqref{eq:txyanyonlatticetrans}. Therefore, lattice transformations do not change the gauge charge carried by $p^{(x)}$ and $p^{(y)}$ and consequentially act on the gauge fields $a^{(5)}$ and $a^{(6)}$ as the identity. Thus, these gauge fields satisfy the typical periodic boundary conditions 
\begin{align*}
    a^{(5)}(x+L_{x},y,t) &= a^{(5)}(x,y+L_{y},t) = a^{(5)}(x,y,t),\\
    a^{(6)}(x+L_{x},y,t) &= a^{(6)}(x,y+L_{y},t) = a^{(6)}(x,y,t).
\end{align*}
Their four holonomies therefore need to go around the system only once to close and are given by
\begin{align*}
    \Gamma_{7}(y,t) &= \oint_{0}^{L_{x}} \mathrm{d}x~a^{(5)}_{x}(x,t,y),\\
    \Gamma_{8}(x,t) &= \oint_{0}^{L_{y}} \mathrm{d}y~a^{(5)}_{y}(x,t,y),\\
    \Gamma_{9}(y,t) &= \oint_{0}^{L_{x}} \mathrm{d}x~a^{(6)}_{x}(x,t,y),\\
    \Gamma_{10}(x,t) &= \oint_{0}^{L_{y}} \mathrm{d}y~a^{(6)}_{y}(x,t,y).
\end{align*}

The other two holonomies from the gauge charge sector can be found by consider a particle carrying $\mathfrak{e}$ gauge charge, which corresponds to the gauge field $a^{(4)}$. However, now the corresponding anyon lattice vector, ${\ell_{e}^{i} = \delta^{i4}}$, transforms non-trivially. Indeed, going around the system $n^{e}_{x}$ times in the $x$-direction or $n^{e}_{y}$ times in the $y$-direction, this anyon lattice vector transform as
\begin{align*}
    \left(T_{x}^{(\mathcal{A})}\right)^{n^{e}_{x}L_{x}}:\quad &\vec{\ell}_{e}\to \vec{\ell}_{e} + (n^{e}_{x}L_{x}~\operatorname{mod}~N)~\vec{\ell}_{p_{x}},\\
    \left(T_{y}^{(\mathcal{A})}\right)^{n^{e}_{y}L_{y}}:\quad &\vec{\ell}_{e}\to \vec{\ell}_{e} + (n^{e}_{y}L_{y}~\operatorname{mod}~N)~\vec{\ell}_{p_{y}}.
\end{align*}
Therefore, in order for the holonomy in terms of the corresponding gauge fields to close, the number of times it winds around the system must satisfy ${n^{e}_{i}L_{i}~\operatorname{mod}~N = 0}$. For example, for the $x$-direction, this implies that
\begin{equation*}
    n^{e}_{x}L_{x} - k N = 0
\end{equation*}
for any integer $k$. The smallest value $k$ can take such that $n_{x}^{e}$ is an integer is ${k = L_{x}/\gcd(L_{x},N)}$ and therefore ${n^{e}_{x} = N/\gcd(L_{x},N)}$. Using that the least-common multiple ($\lcm$) between two integers $a$ and $b$ satisfies ${\lcm(a,b) = |ab|/\gcd(a,b)}$, the gauge field $a^{(4)}$ satisfies the boundary condition
\begin{equation*}
     a^{(4)}(x+\lcm(L_{x},N),y,t) = a^{(4)}(x,y,t).
\end{equation*}
Similarly, in $y$ direction $a^{(4)}$ must also satisfy the boundary condition
\begin{equation*}
     a^{(4)}(x,y+\lcm(L_{y},N),t) = a^{(4)}(x,y,t).
\end{equation*}
Therefore, the two holonomies involving $a^{(4)}$ are
\begin{align*}
    \Gamma_{11}(y,t) &= \oint_{0}^{\lcm(L_{x},N)}\mathrm{d}x~a^{(4)}_{x}(x,y,t),\\
    \Gamma_{12}(x,t) &= \oint_{0}^{\lcm(L_{y},N)}\mathrm{d}y~a^{(4)}_{y}(x,y,t).
\end{align*}

The other six holonomies come from the gauge flux sector. Similar to the $p^{(x)}$ and $p^{(y)}$ excitations in the gauge charge sector, the lattice vector for the $g$ excitation does not change under lattice translations. Therefore, the corresponding gauge field, $a^{(3)}$, satisfies normal periodic boundary conditions
\begin{equation*}
    a^{(3)}(x+L_{x},y,t) = a^{(3)}(x,y+L_{y},t) = a^{(3)}(x,y,t),
\end{equation*}
from which we introduce the two holonomies
\begin{align*}
    \Gamma_{1}(y,t) &= \oint_{0}^{L_{x}} \mathrm{d}x~a^{(3)}_{x}(x,t,y),\\
    \Gamma_{2}(x,t) &= \oint_{0}^{L_{y}} \mathrm{d}y~a^{(3)}_{y}(x,t,y).
\end{align*}
Similarly, acting the lattice translations on the anyon lattice vectors of $m^{x}$ and $m^{y}$ in their longitudinal directions leaves them unchanged. And so, their corresponding gauge fields satisfy periodic boundary conditions in the $x$ and $y$-directions, respectively:
\begin{align*}
    a^{(1)}(x+L_{x},y,t) &= a^{(1)}(x,y,t),\\
    a^{(2)}(x,y+L_{y},t) &= a^{(2)}(x,y,t).
\end{align*}
Their corresponding holonomies are
\begin{align*}
    \Gamma_{3}(x,t) &= \oint_{0}^{L_{y}} \mathrm{d}y~a^{(2)}_{y}(x,t,y),\\
    \Gamma_{4}(y,t) &= \oint_{0}^{L_{x}} \mathrm{d}x~a^{(1)}_{x}(x,t,y).
\end{align*}

The last two holonomies come from $m^{(x)}$ and $m^{(y)}$ going around the system in their transverse directions. Consider going around the system $n^{m}_{x}$ or $n^{m}_{y}$ times in both transverse directions:
\begin{align*}
    \left(T_{y}^{(\mathcal{A})}\right)^{n^{m}_{y}L_{y}}:\quad& \vec{\ell}_{m^{(x)}}\to \vec{\ell}_{m^{(x)}} + (n^{m}_{y}L_{y}~\operatorname{mod}~N)~\vec{\ell}_{g},\\
    \left(T_{x}^{(\mathcal{A})}\right)^{n^{m}_{x}L_{x}}:\quad& \vec{\ell}_{m^{(y)}}\to \vec{\ell}_{m^{(y)}} + (-n^{m}_{x}L_{x}~\operatorname{mod}~N)~\vec{\ell}_{g}.
\end{align*}
Notice that both transformations add $\vec{\ell}_{g}$. Therefore, the most general holonomy will include hopping both $m^{(x)}$ in the $y$-direction and $m^{(y)}$ in the $x$-direction. Indeed, letting
\begin{align*}
    \vec{\ell'}_{m^{(x)}} &= \left(T_{y}^{(\mathcal{A})}\right)^{n^{m}_{y}L_{y}}\vec{\ell}_{m^{(x)}},\\
    \vec{\ell'}_{m^{(y)}} &= \left(T_{x}^{(\mathcal{A})}\right)^{n^{m}_{x}L_{x}}\vec{\ell}_{m^{(y)}},
\end{align*}
we have that
\begin{equation*}
     \vec{\ell'}_{m^{(x)}} +\vec{\ell'}_{m^{(y)}} \hspace{-2pt}= \vec{\ell}_{m^{(x)}} +\vec{\ell}_{m^{(y)}} + (n^{m}_{y}L_{y}-n^{m}_{x}L_{x}~\operatorname{mod}~N))\vec{\ell}_{g}.
\end{equation*}
So, in order for this holonomy to close, we must have $n_{y}^{m}$ and $n_{x}^{m}$ satisfy ${n^{m}_{y}L_{y}-n^{m}_{x}L_{x}~\operatorname{mod}~N = 0}$, which implies that for some integer $k$ that
\begin{equation}\label{eqn:twoDimHolCons}
    n^{m}_{y}L_{y}-n^{m}_{x}L_{x}-kN = 0.
\end{equation}

We are looking for two holonomies from which all other holonomies involving $m^{(x)}$ and $m^{(y)}$ moving in their transverse directions can be generated. They are defined by two different ``basis'' values of $n_{y}^{m}$ and $n_{x}^{m}$: $(n_{x}^{m1}, n_{y}^{m1})$ and $(n_{x}^{m2}, n_{y}^{m2})$. For the first pair, $(n_{x}^{m1}, n_{y}^{m1})$, we are free to have one of the components be zero, for instance ${n_{x}^{m1} = 0}$. Then, this reduces to the one dimensional version we've considered previously, so ${(n_{x}^{m1}, n_{y}^{m1}) = (0,N/\gcd(L_{y},N))}$. This then gives the holonomy.
\begin{equation*}
   \Gamma_{5}(x,t) = \oint_{0}^{\lcm(L_{y},N)}\mathrm{d}y~a^{(1)}_{y}(x,y,t). 
\end{equation*}
Because $n_{x}^{m1} = 0$, $n_{x}^{m2}$ must be the smallest possible value to ensure all holonomies can be generated. Rewriting Eq.~\eqref{eqn:twoDimHolCons} as
\begin{equation*}
    n^{m}_{x}L_{x}-\gcd(L_{y},N)\left(n^{m}_{y}\frac{L_{y}}{\gcd(L_{y},N)}-k\frac{N}{\gcd(L_{y},N)}\right) = 0,
\end{equation*}
the term in parenthesis is some integer in terms of the variables we're solving for, and so like before the smallest $n^{m}_{x}$ that satisfies this is 
\begin{equation*}
    n_{x}^{m2} = \frac{\gcd(L_{y},N)}{\gcd(L_{x},\gcd(L_{y},N))} = \frac{\gcd(L_{y},N)}{\gcd(L_{x},L_{y},N)}.
\end{equation*}
Plugging this back into Eq.~\eqref{eqn:twoDimHolCons}, $n^{m2}_{y}$ is given by
\begin{equation}
    n^{m2}_{y} = \frac{\lcm(L_{x},\gcd(L_{y},N))+kN}{L_{y}},
\end{equation}
where $k$ is any integer for which $n^{m2}_{y}$ is also an integer. There does not appear to be a closed form for such a $k$ in terms of generic $L_{x}$, $L_{y}$, and $N$. Nevertheless, from the theory of linear Diophantine equations, because $\gcd(L_{y},N)$ divides $\lcm(L_{x},\gcd(L_{y},N))$ there indeed exists a solution~\cite{jones2012elementary}. Therefore, from here on out we will leave our expressions in terms of the integer $n_{y}^{m2}$. From the above discussion we therefore have that the final holonomy is given by
\begin{align*}
    \Gamma_{6}(x,y,t) &= \oint_{0}^{\lcm(L_{x},\gcd(L_{y},N))}\mathrm{d}x~a^{(2)}_{x}(x,y,t)\\
    &\hspace{40pt}+ \oint_{0}^{n^{m2}_{y}L_{y}}\mathrm{d}y~a^{(1)}_{y}(x,y,t).
\end{align*}

\subsection{Effective Action and Ground State Degeneracy}\label{sec:effectiveAction}

Having written down the Mutual Chern Simons theory and the corresponding Holonomies of the torus, we can now find an effective theory describing the degenerate ground state manifold on the rank-2 toric code. The low-energy local constraint defining the ground state $\ket{\text{vac}}$ of the mutual Chern-Simons theory is given by the Gauss-law constraint\footnote{By Gauss-law constraints, we mean the local constraints which the $0$-components of $a$ act as Lagrange multipliers to enforce. Or, equivalently for the mutual Chern-Simons theory, the equations of motion for $a^{(i)}_{0}$.} ${K_{0i}~\epsilon^{jk}\partial_{j}a^{(i)}_{k}\ket{\text{vac}} = 0}$, which implies that ${\epsilon^{0jk}\partial_{j}a^{(l)}_{k}\ket{\text{vac}} = 0}$. We note that upon quantizing the theory, the operator $\epsilon^{0jk}\partial_{j}a^{(l)}_{k}$ is the generator the familiar $U(1)$ gauge transformation ${a^{(i)}_{k}\to a^{(i)}_{k}+\partial_{k} f^{(i)}}$. Considering only states within the low-energy subspace of the Hilbert space, the gauge fields satisfy $\partial_{x}a^{(i)}_{y} = \partial_{y}a^{(i)}_{x}$ at all points in spacetime. This causes constraints to arise on the gauge-invariant line-integrals $\Gamma_{i}$. For instance, consider $\Gamma_{1}(y,t)$ and differentiate it with respect to $y$. Pulling the partial derivative inside the integral and using the Gauss law constraint, we find that $\partial_{y}\Gamma_{1}(y,t) = 0$. Using similar manipulations, it's easy to show that all $\Gamma_{i}$ are position independent and only depend on $t$. Then, under the influence of this constraint, the 12 holonomies of the torus are constrained to $\Gamma_{i}(x,y,t) = \varphi_{i}(t)$.

Because this is true for any gauge-field configuration, we can express the components gauge fields in terms of these space-independent 12 holonomies. For the holonomies integrating around space only once ($\Gamma_{i}$ for $i=1,2,3,4,7,8,9,10$) this give the familiar results~\cite{wen2004quantum}
\begin{align*}
    a^{(3)}_{x}(x,y,t) &= \frac{\varphi_{1}(t)}{L_{x}},
    \hspace{40pt} a^{(3)}_{y}(x,y,t) = \frac{\varphi_{2}(t)}{L_{y}},\\
    a^{(5)}_{x}(x,y,t) &= \frac{\varphi_{7}(t)}{L_{x}},
    \hspace{40pt} a^{(5)}_{y}(x,y,t) = \frac{\varphi_{8}(t)}{L_{y}},\\
    a^{(6)}_{x}(x,y,t) &= \frac{\varphi_{9}(t)}{L_{x}},
    \hspace{40pt} a^{(6)}_{y}(x,y,t) = \frac{\varphi_{10}(t)}{L_{y}},\\
    a^{(2)}_{y}(x,y,t) &= \frac{\varphi_{3}(t)}{L_{x}},
    \hspace{40pt} a^{(1)}_{x}(x,y,t) = \frac{\varphi_{4}(t)}{L_{y}}.
\end{align*}
For the holonomies that can wind around the system multiple times ($\Gamma_{i}$ for $i = 5,6,11,12$), we find that 
\begin{align*}
    a^{(4)}_{x}(x,y,t) &= \frac{\varphi_{11}(t)}{\lcm(L_{x},N)},
    \hspace{10pt} a^{(4)}_{y}(x,y,t) = \frac{\varphi_{12}(t)}{\lcm(L_{y},N)},\\
    a^{(1)}_{y}(x,y,t) &= \frac{\varphi_{5}(t)}{\lcm(L_{y},N)},\\
    a^{(2)}_{x}(x,y,t) &= \frac{N~\varphi_{6}(t) - n^{m2}_{y} \gcd(L_{y},N)~\varphi_{5}(t) }{N\lcm(L_{x},\gcd(L_{y},N))}.
\end{align*}
These expressions for the gauge fields in terms of the holonomies are defined up to some pure gauge fluctuations, which we do not include. The part we do show is independent of space, giving the topological part of the gauge fields and acting as their zero-momentum modes. Lastly, we emphasize that because in a compact gauge theory the low-energy observables are Wilson loop amplitudes $W_{i} = \mathrm{e}^{\mathrm{i}\varphi_{i}}$, the holonomies $\varphi_{i}$ are all $2\pi$ periodic phases.

Plugging in this expressions to the mutual Chern-Simons theory, we get an effective theory of ground state in terms of the holonomies (zero modes of gauge fields) described by the action
\begin{equation}\label{eqn:Seffbij}
    S_{\text{eff}} = \frac{b^{ij}}{2\pi}\int \mathrm{d}t~ \varphi_{i}\frac{\mathrm{d}\varphi_{j}}{\mathrm{d} t}.
\end{equation}
Here, $b$ is a 12 $\times$ 12 antisymmetric matrix given by
\begin{equation}
    b = 
    \begin{pmatrix}
    0_{6} & B\\
    -B^{\top} & 0_{6}
    \end{pmatrix},
\end{equation}
where $0_{6}$ is a $6\times 6$ matrix of zeros and the integer matrix $B$ is
\begin{equation}
    B = 
    \begin{pmatrix}
    0 & 0                  & 0     & 0  & 0      & N_{y} \\
    0 & 0                  & 0     & 0  & -N_{x} & 0     \\
   -N & 0                  & 0     & 0  & 0      & 0     \\
    0 & 0                  & 0     & -N & 0      & 0     \\
    0 & -n^{m2}_{y}N_{xy}       & N_{y} & 0  & 0      & 0     \\
    0 & N N_{xy}N_{y}^{-1} & 0     & 0  & 0      & 0     \\
    \end{pmatrix}.
\end{equation}
For conciseness in the matrix $B$ we have denoted
\begin{align*}
    N_{x} &\equiv \gcd(L_{x},N),\\
    N_{y} &\equiv \gcd(L_{y},N),\\
    N_{xy} &\equiv \gcd(L_{x},L_{y},N).
\end{align*}
$S_{\text{eff}}$ describes a particle moving on a 12-torus in a magnetic field described by the two form $b_{ij}$.

Using the effective theory, we can now see if it reproduces the ground state degeneracy found from the anyon lattice calculation in Section~\ref{sec:anyonLatticeGSD}. Because the Wilson loop amplitudes are the only observables at low-energy, the number of ground states is given by size of the smallest faithful representation of the nontrivial commutation relations satisfied by $W_{i}$. Quantizing the effective theory, from the canonical commutation relation ${\left[b^{ij}\varphi_{i}/2\pi,\varphi_{j}\right] = \mathrm{i}}$ (where $i$ is summed over but $j$ is not), we find that the commutation relations between holonomies is ${[\varphi_{i},\varphi_{j}] = 2\pi\mathrm{i} (b^{-1})_{ij}}$. From this, the algebra satisfied by Wilson loops operators is therefore given by
\begin{equation*}
    W_{i}W_{j} = \mathrm{e}^{-2\pi\mathrm{i}(b^{-1})_{ij}}W_{j}W_{i}.
\end{equation*}
The dimension of the smallest representation is given by the Pfaffian of $b$~\cite{wesolowski1994multiple}, and therefore the ground state degeneracy is ${\text{GSD} = |\operatorname{pf}(b)|}$. Computing the Pfaffian, we find that
\begin{equation}
\text{GSD} = N^{3}\gcd(L_{x},N)\gcd(L_{y},N)\gcd(L_{x},L_{y},N),
\end{equation}
exactly agreeing with the result Eq.~\eqref{eqn:GSD} found from the anyon lattice. Note that despite the variable $n^{m2}_{y}$ appearing in the matrix $B$, it does not affect the GSD.

The action $S_{\text{eff}}$ of Eq.~\eqref{eqn:Seffbij} describes the long-wavelength properties of the SET order in the rank-2 toric code, yet the number of ground states depends on details from the lattice (the number of lattice sites, $L_{x}$ and $L_{y}$). This arises because the effective action's coupling constants are explicitly dependent on these microscopic parameters, and thus we see the UV/IR mixing directly. 
The structure of these coupling constants are a consequence of the structure of the holonomies found in Section~\ref{sec:holonomies}.
Therefore, from the point of view of field theory, the UV/IR mixing is due to the twisted boundary conditions satisfied by the gauge fields, which forced the holonomies to integrate around the system a number of times dependent on the microscopic parameters in order to be gauge invariant. And finally, similar to the understanding of UV/IR mixing from the anyon lattice point of view, this was determined by how lattice translations act on anyon lattice vectors, and hence the gauge fields. Therefore, microscopically, the UV/IR mixing emerges due to the interplay between the lattice's translation symmetries and the long-range entanglement pattern in the many body states of the rank-2 toric code.

\section{Conclusion}

In this paper, we have built off the work initiated in Refs.~\cite{PhysRevB.97.235112, PhysRevB.98.035111,oh2021path} and studied an exactly solvable point of a Higgsed symmetric tensor gauge theory in $2+1$d known as the $\mathbb{Z}_{N}$ rank-2 Toric code. This model has unconventional symmetry enriched topological (SET) order, meaning that the enriched symmetries permute inequivalent anyon types in addition to acting on them projectively. We found that this enforces anyons of the same species to have a spatially dependent flavor index based on the different gauge charge/flux they carry. Using this, we investigated their mobility, position-dependent braiding statistics, and how the lattice transformations are realized on the anyon lattice. This allowed us to find the ground state degeneracy on a torus for general $N$, which revealed the presence of UV/IR mixing. Then, using the basis charges and fluxes of the anyon lattice, we developed a mutual Chern-Simons theory from which we found a low-energy effective action describing the SET order of the rank-2 toric code. This low energy theory on a torus reproduced the ground state degeneracy and explicitly showed the presence of microscopic details in its coupling constants, and hence the aforementioned UV/IR mixing.

There are many interesting follow-up questions. The first few are in the context of the rank-2 Toric code model. It is a very rich model due to how the lattice symmetries couple to the topological order, and it would be interesting to explore their interplay further. For instance, the symmetry fractionalization patterns~\cite{chen2017symmetry} arising could be rich and exciting to understand. Furthermore, upon condensing anyons, since the anyons are position-dependent, not only would the topological order change~\cite{PhysRevB.80.125101} but some of the lattice symmetries would also spontaneously break. Studying these lattice-symmetry breaking patterns from condensing topological excitations would be interesting too. Additionally, because excitations have directions which they can only hop by greater than one lattice site, it would also be interesting to study the effect of extrinsic lattice defects, which would act as non-abelian excitations~\cite{PhysRevLett.105.030403,kitaev2012models, PhysRevB.86.161107, PhysRevX.2.031013, PhysRevB.87.045130, 10.21468/SciPostPhys.12.2.064}. 

The second set of follow up questions are in the context of fracton topological order. The unconventional SET order in the rank-2 Toric code model is reminiscent of fracton topological order. Therefore, it would be interesting to see if the ideas presented here could be applied to a model with genuine fracton topological order. For example, could emergent conservation laws arising from the fusion rules be used to find the position-dependent gauge charge/flux in a model with fracton topological order, as we did in Section~\ref{sec:anyonLattice} for the rank-2 toric code? In doing so, we found that despite there being $N^{2}+2N$ inequivalent elementary excitations, there were fewer types of gauge charge and flux: only six basis charges/fluxes. A model with fracton topological order would start off with an extensive number of inequivalent elementary excitations, but its likely that by using the emergent conservation laws, one would end up with a basis including a sub-extensive number of charges/fluxes. From this basis, assuming the results from conventional topological order apply, it becomes quite obvious that the ground state degeneracy should scale sub-extensively with the system's size. Furthermore, it's an open question for whether or not there exists an effective quantum field theory that describes the low-energy physics of fracton topological order~\cite{PhysRevB.96.195139, PhysRevResearch.2.023249,radicevic2019systematic,bulmash2018generalized, 10.21468/SciPostPhysCore.4.2.012,10.21468/SciPostPhys.6.4.043, PhysRevX.9.031035,PhysRevB.103.195113, gorantla2022global}. From the position-dependent excitations point of view, there would be a sub-extensive number of corresponding Chern-Simons gauge fields, which is closely in line with the thinking of the infinite Chern-Simons theory developed in Ref.~\cite{ma2020fractonic}. Additionally, such a framework could open up the avenue to consider fracton topological order in terms of $S$ and $T$ matrix formalism~\cite{wen2016theory}, which could possibly be studied in the thermodynamic limit in which they would be infinitely dimensional matrices. However, how one could extract the mutual statics data required remains an open question.

\section{Acknowledgements}

S.D.P. is grateful to Claudio Chamon, Arkya Chatterjee, Michael DeMarco, Ethan
Lake, Aidan Reddy, and Guilherme Delfino Silva for fun and interesting
discussions. Furthermore, S.D.P. acknowledges support from the Henry W. Kendall
Fellowship.  This work is partially supported by NSF DMR-2022428 and by the
Simons Collaboration on Ultra-Quantum Matter, which is a grant from the Simons
Foundation (651446, XGW).

\bibliographystyle{apsrev4-1}
\bibliography{references} 

\end{document}